\documentclass[paper,11pt]{article}
\pdfoutput=1

\usepackage{jcappub}

\newcommand{\Darkside}{{DarkSide}}
\newcommand{\darkside}{{DarkSide }}

\definecolor{agrey}{rgb}{0.1000,0.4000,0.1000}
\definecolor{ared}{rgb}{0.7000,0.1000,0.1000}

\title{A Fluka Study of Underground Cosmogenic Neutron Production}

\author[1]{A. Empl,\note{Corresponding author.}}
\author[]{E. V. Hungerford,}
\author[]{R. Jasim}
\author[b]{and P. Mosteiro}

\affiliation[]{Department of Physics, University of Houston, Houston, TX 77204}
\affiliation[b]{Department of Physics, Princeton University, Princeton, NJ 08544}

\emailAdd{aempl@central.uh.edu}

\abstract{
 Neutrons produced by cosmic muon interactions are important contributors
 to backgrounds in underground detectors when searching for rare events.
 Typically such neutrons can dominate the background, as they are
 particularly difficult to shield and detect.  Since actual data is
 sparse and not well documented, simulation studies must be used to
 design shields and predict background rates. Thus validation of
 any simulation code is necessary to assure reliable results. This work
 compares in detail predictions of the FLUKA simulation code to existing
 data, and uses this code to report a simulation of cosmogenic backgrounds
 for typical detectors embedded in a water tank with liquid scintillator
 shielding.
}

 \keywords{Muon, Cosmic, Cosmogenic, Neutron, Borexino, DarkSide}

\begin{document}
\maketitle
\flushbottom

\section{Introduction}

 This paper reports the use of the FLUKA simulation code to study
 muon-induced, cosmogenic backgrounds.  It is
 particularly interested in predicted neutron backgrounds for direct 
 neutrino and dark matter experiments, and additionally in the production
 of radio-isotopes in the detector shielding.  Typically,
 such experiments are placed in deep underground facilities where
 only neutrinos and high energy muons are able to reach the detectors.
 In this study, special care was taken to validate FLUKA with available
 data, and by comparing our results with previous simulations.
 Muons induce backgrounds by producing showers of secondary
 particles in local interactions in, or near, the detector. As only
 high energy muons penetrate to the depth of the experimental
 halls, the total muon flux decreases as the lower energy, more
 intense muons, are removed from the spectrum.  Consequently, the mean
 energy of the residual muons increases with depth.

 Little experimental information about muon-induced secondaries at
 depth is available.  Most previous studies concentrated on the neutron
 flux in the muon radiation field.  Low rates and challenges inherent
 in neutron measurements require careful interpretation of the available
 data.  Indeed, measurements of neutron production at deep underground
 sites was previously suggested as a method to study the incident muon
 flux at such facilities ~\cite{zat65}.  As a consequence, neutron yield
 was measured at increasing depth for  many years
 \cite{bezrukov0}-\nocite{enikeev,lvdAgli,enik2,lsdAgli}\cite{boehm}.
 However, such information was limited to cosmogenic neutrons, mostly
 in large, liquid-scintillator detectors, where the
 flux was determined by measuring the emitted gamma radiation
 after neutron capture in the scintillator. Thus detector
 geometry and efficiency affected the interpretation of the data, and
 experiments had to rely on the simulation of
 particle interactions and transport to extract and predict the
 muon-induced backgrounds. Interpretation of the early measurements
 was (and is) challenging, since systematic  uncertainties are large
 and usually underestimated.

\section{FLUKA}

\subsection{The FLUKA Code}

 FLUKA~\cite{fluka1, fluka2} is a fully integrated particle-physics,
 Monte Carlo simulation package, originally introduced to aid in
 shielding design of particle accelerators.  More recently it has been applied
 in high energy particle physics, medical physics, radio-biology, 
 and of relevance here, to simulate cosmic ray and cosmogenic
 backgrounds in deep underground experiments.    

 Design and development of FLUKA has always been based on the
 implementation of verified microscopic models of physical processes.
 FLUKA uses these models in a way
 which maintains consistency among all the reaction steps and
 types.  Thus, all conservation laws are enforced at each step and predictions
 are benchmarked against experimental data, if possible, at the level of single     
 interactions.   This results in a consistent approach to all
 energy/target/projectile combinations with a minimal overall
 set of free parameters.   As a consequence, predictions for complex
 simulation problems are robust and arise naturally from
 underlying physical models. Depending on model validity, FLUKA is
 expected to provide reasonable results even in cases where no direct 
 experimental data are available~\cite{manual}.

 The version of FLUKA used for the present study is FLUKA2011.2,
 from November 2011.  A collection of benchmark results for the physical
 aspects of the code relevant to the problem of muon-induced backgrounds
 are presented in \cite{us}.

 \subsection{Physics models in FLUKA}
 \label{fluka.models}

 Production of cosmogenic neutrons in FLUKA is the result of direct
 muon nuclear interactions, photo-nuclear reactions by real photons in 
 electromagnetic showers, and in nuclear cascades within resulting
 hadronic showers. Direct muon-nuclear interactions are modeled by
 $\mu^{-}$ capture at rest, and by virtual photo-nuclear interactions.
 The latter are factorized following Bezrukov-Bugaev~\cite{bezrukov}
 into virtual photon production and photon-nucleus reactions.
 The photon-nucleus reactions are simulated over the entire energy range
 through different mechanisms:
  \begin{itemize}
  \item  Giant Resonances interaction,
  \item  Quasi-Deuteron effect,
  \item  Delta Resonances production, and
  \item  Vector Meson Dominance at high energies.
  \end{itemize}

 Hadronic interactions in FLUKA are described in several papers
 (\cite{bib:fluka4}\nocite{bib:fluka5}\nocite{bib:fluka6}-\cite{bib:fluka7}).
 Hadron-nucleon inelastic collisions up to a few GeV are realized                   
 in terms of resonance production and decay.  At higher energies, the
 simulation 
 employs a  modified  Dual Parton Model~\cite{bib:fluka7}.  The Dual
 Parton Model is a quark/parton string model providing
 reliable results up to several tens of TeV.

 The FLUKA nuclear interaction model called PEANUT [3­6] can be schematically
 described as a sequence of the following steps: 
  \begin{itemize}
  \item  Glauber-Gribov cascade in high energy collisions,
  \item  Generalized-Intra-Nuclear cascade,
  \item  Pre-equilibrium emission, and
  \item  Evaporation/Frag\-mentation/Fission and de-excitation.
  \end{itemize}
 Some of these steps may be skipped depending on the projectile energy and
 type.  PEANUT is a precise and reliable tool for intermediate energy
 hadron-nucleus reactions.  Its ``nuclear environment'' is also used in
 the simulation of real and virtual photo-nuclear reactions, neutrino
 interactions, nucleon decays and muon captures.
 All nuclear interaction models, including nucleus-nucleus interactions, 
 share parts of the common PEANUT framework.
 In particular, all nuclear fragments, irrespective of how they are produced,
 are de-excited through a common evaporation/fragmentation and
 gamma production chain.  A validation of the FLUKA Monte Carlo code for
 predicting induced radioactivity is given in~\cite{bib:fluka_activation}.

 \subsection{Physics related user options}

 The physics models in FLUKA are fully integrated into the code, and the
 individual models are benchmarked against available experimental data.     
 The user is presented with an overall
 optimized configuration of models which cannot be user adjusted.

 The simulation reported here was performed with the FLUKA default setting
 PRECISIO(n).   In addition, photo-nuclear interactions were enabled through
 the FLUKA option PHOTONUC and the detailed treatment of nuclear
 de-excitation was requested with the EVAPORAT(ion) and COALESCE(nce)
 options.   The latter two options are suggested to the user in order to
 obtain more reliable results for isotope production.   These enable the
 evaporation of heavy fragments (A$>$1) and the emission of energetic
 light-fragments, respectively.  The treatment of nucleus-nucleus
 interactions was also turned on for all energies via the option IONTRANS,
 and delayed reactions were enabled through the option, RADDECAY.

 Neutron captures on hydrogen inside liquid
 scintillator are recorded in order to evaluate the muon-induced neutron
 production rate. This approach closely follows that of any such
 experimental measurement and avoids technical ambiguities for neutron
 counting.


\section{Cosmogenic background simulation at LNGS}

 A faithful simulation of the muon radiation field in the vicinity of
 any underground experiment requires detailed information of the depth,
 overburden geometry, and composition of the rock surrounding the detector
 through which the muons propagate.  Since the
 cavern size is small relative to changes in the flux, the
 radiation field can be assumed constant at the average depth of the
 experimental hall.  However, details
 of the detector geometry and materials surrounding the cavern and detector
 must be included in any accurate simulation.

 This study differs from previous work in that it includes multi-muon events.
 As will be seen later, this improves the description of the
 radiation field.
   A common approach found in previous literature in context of
   predicting the neutron yield in liquid scintillator makes use
   of mono-energetic muons
 with a mean kinetic energy corresponding to the average depth of the
 underground site.  This study finds that the predicted neutron yield
 decreases by approximately 8\% if mono-energetic muons of 280 GeV are
 replaced by muons of an appropriate differential energy spectrum having
 the same mean value for the kinetic energy.  However, if muon bundles
 are properly included, an overall increase in neutron yield on the order
 of 4\% results.


 \subsection{Muon radiation field}

 \subsubsection{Intensity}
 
 The muon flux in the 3 halls at the Italian
                         {\it Laboratori Natzionali del Gran Sasso}
 (LNGS) for Underground research is summarized in a recent
 publication~\cite{seasonal}.
 More precisely, the measured flux describes the total cosmogenic
 muon event rate, since individual events may contain more than one
 muon~\cite{decoherence}.
 Variations in muon flux between the
 measurements given in Table~\ref{mu_flux} may be attributed to the
 relative location of the halls within the laboratory, differences in time
 periods of data collection, and perhaps more importantly,
 the systematic uncertainties.  The value of the flux given by Borexino
 located in Hall C was adopted for the present simulation.

\begin{table}[htb]
 \begin{center}
   \begin{tabular}{lccl}
    experiment & Hall & year published & total muon event rate \\
               &      &                & ($\times 10^{-4} \,\, s^{-1}
               m^{-2}$) \\ 
    \hline
    LVD      & A & 2009 & ${\bf 3.31} \pm {\bf 0.03}$ \\
    MACRO    & B & 2002 & ${\bf 3.22} \pm {\bf 0.08}$ \\
    Borexino & C & 2012 & ${\bf 3.41} \pm {\bf 0.01}$ \\
   \end{tabular}
   \caption{The table shows the measured muon flux in the
              various experimental halls at the LNGS}
   \label{mu_flux}
\end{center}
\end{table}

\subsubsection{Mean energy and differential energy spectrum}

 The underground muon kinetic energy spectrum can be represented
 by the following (for example \cite{macro02}):

 $$
    {dN\over{dE_{\mu}}} = const \cdot (E_{\mu} + \epsilon(1-e^{-\beta
    h}))^{-\alpha} 
   \label{eq:one}
 $$

 \hskip -\parindent
 In the above equation,  $E_{\mu}$  is the muon kinetic energy at slant depth
 $h$, and $\alpha$ is the surface muon spectral index, and the
 quantities $\beta$ and $\epsilon$ are related to 
 muon energy loss mechanisms in rock.
 The average muon kinetic energy at slant
 depth $h$ is:

 $$
    <\!E_{\mu}\!> \, = \, {\epsilon(1-e^{-\beta h})\over{\alpha -2}}
 $$

 \vskip 1mm \hskip -\parindent
 Experimental results for the spectral index $\alpha$ and the mean
 muon energy for LNGS were reported by MACRO \cite{macro02} and are
 reproduced in Table~\ref{index}:

   \begin{table}[hbt]
    \begin{center}
   \begin{tabular}{lcl}
     event type  & mean muon energy (GeV)                          &
     spectral index $\alpha$ \\ 
     \hline
     single muon & 270 $\pm$ \, 3 $_{(stat)}$ $\pm$ 18 $_{(syst)}$ &
     3.79 $\pm$ 0.02 $_{(stat)}$ $\pm$ 0.11 $_{(syst)}$ \\ 
     double muon & 381 $\pm$ 13 $_{(stat)}$ $\pm$ 21 $_{(syst)}$   &
     3.25 $\pm$ 0.06 $_{(stat)}$ $\pm$ 0.07 $_{(syst)}$ \\ 
     \end{tabular}
     \caption{The mean energy of single and double muon events
               as measured by MACRO}
     \end{center}
   \label{index}
   \end{table}

 \hskip -\parindent
 The functional description of the energy spectrum given by equation
 (\ref{eq:one}) permits direct sampling since it can be integrated
 and inverted in analytic form.  Adopting the values 
 $\epsilon = 0.392 \times 10^{-3}$ and $\beta = 635$~GeV this
 procedure reproduces the measured mean energy for single and double
 muon events respectively, and yields an overall mean residual
 energy of 283\,$\pm$\,19 GeV for cosmogenic muons at LNGS.

 \subsubsection{Angular distribution}

 The slant depth of the muon transit through the rock overburden depends
 on both azimuthal and zenith angles. The azimuthal angular dependence
 of the slant depth in rock, $h \,\, (g \, cm^{-2})$ at LNGS,
 depends on the profile
 of the Gran Sasso mountain covering the experimental cavern.   Hence,
 both the intensity and energy profiles of the muons are a function of
 their incident direction on the detector.

 \begin{figure}[thb]
   \centering
   \includegraphics[width=.495\textwidth]{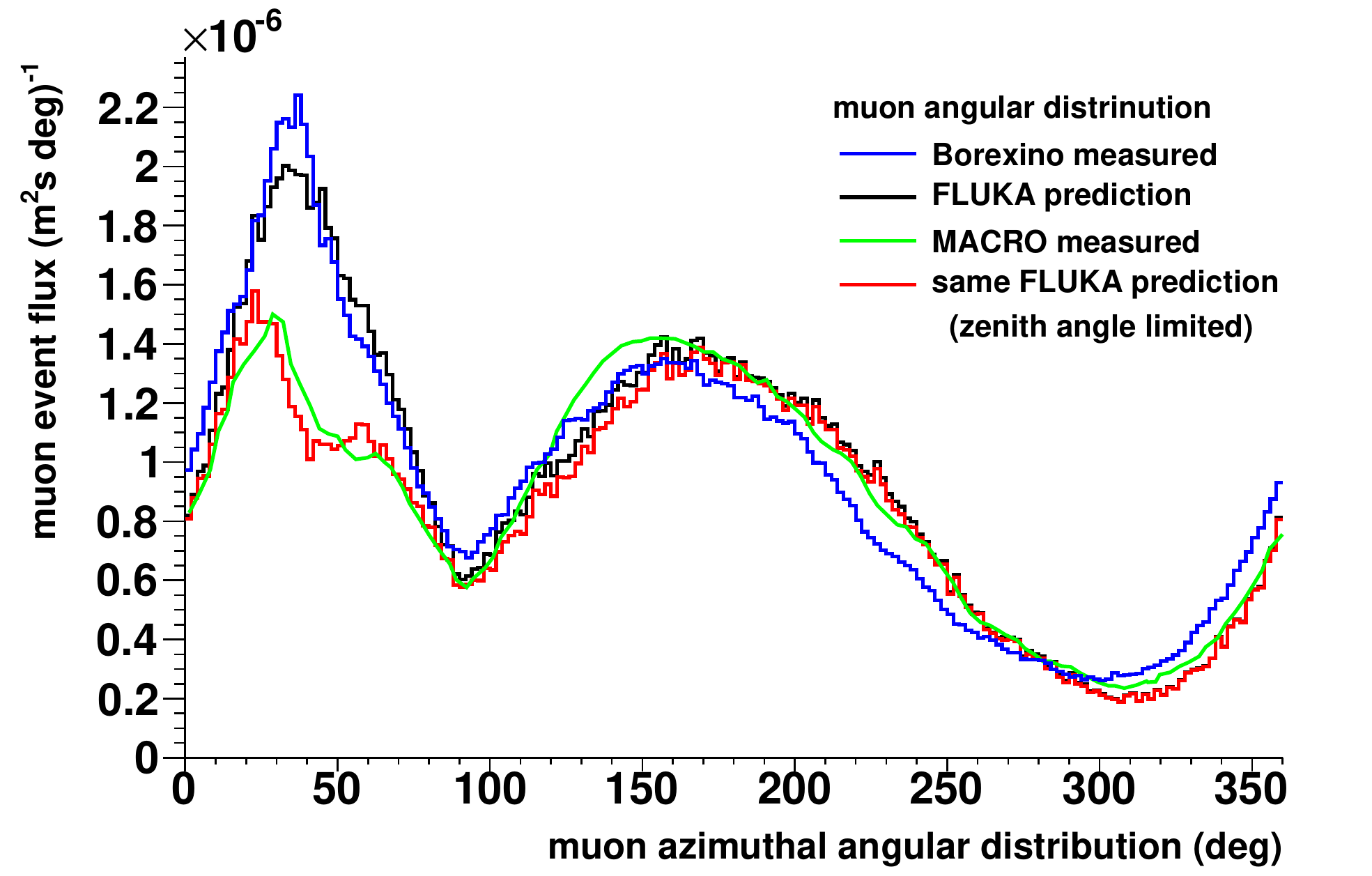}
   \includegraphics[width=.495\textwidth]{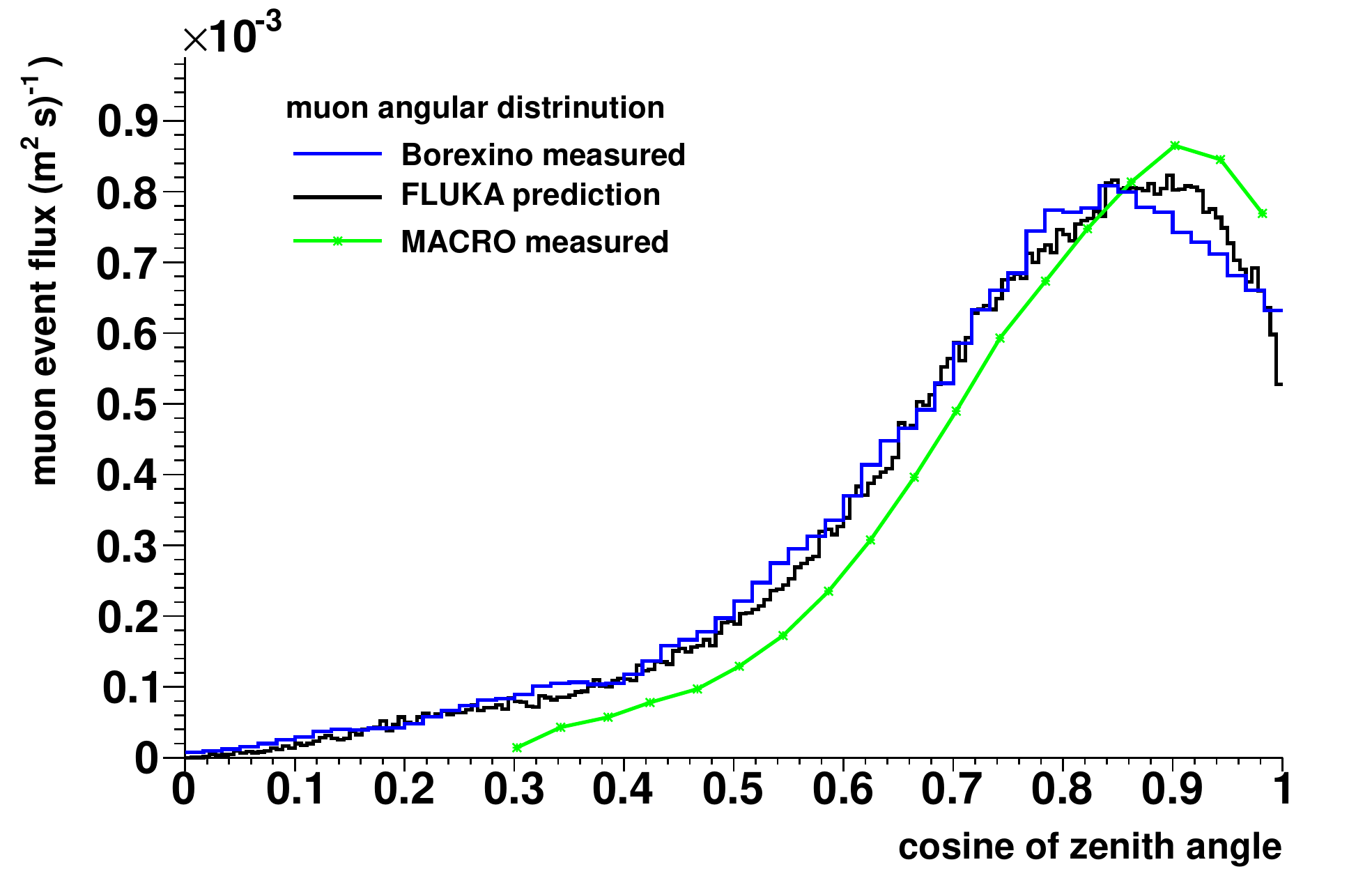}
   \caption{Muon azimuthal (left) and zenith (right) angular
             distribution at LNGS for polar coordinate system pointing
             up and North with clockwise increasing angle.
             Blue line: Borexino data,  green line: MACRO data,
             black (red) line: FLUKA predictions (zenith angle limited).
            }
   \label{fig:angular}
   \end{figure}

 A measurement of the cosmogenic muon angular distribution at LNGS was
 performed by MACRO~\cite{macro93} for Hall B.  Azimuthal (left) and
 zenith (right) projections of the distribution are shown by the green
 histograms in Figure~\ref{fig:angular}. 
 The distributions are compared to recent results from
 Borexino~\cite{cosmogenic_bx} given by the blue histograms.
 The difference in the experimental azimuthal spectra for angles near
 45 degrees is due to an angular acceptance limit in the zenith angle
 of approximately 60 degree in MACRO.  These data are compared
 to a FLUKA simulation which
 traced muons, initiated by cosmic rays in the upper atmosphere, to the
 experimental halls~\cite{max}, and was normalized to the
 Borexino measured total muon flux.   The predictions for full detector
 acceptance are shown by the black histograms in Figure~\ref{fig:angular}.
 When the limitation of zenith angle in MACRO is imposed in FLUKA, 
 the same features found in the data are reproduced.  The resulting
 azimuthal spectrum is shown by the red histogram.  Good agreement is
 found between data and simulations, and the small visible shift in the
 azimuthal distribution is due to the change of location for the two
 experiments.

\subsubsection{Bundles}

\begin{figure}[bht]
    \centering
    \includegraphics[width=.495\textwidth]{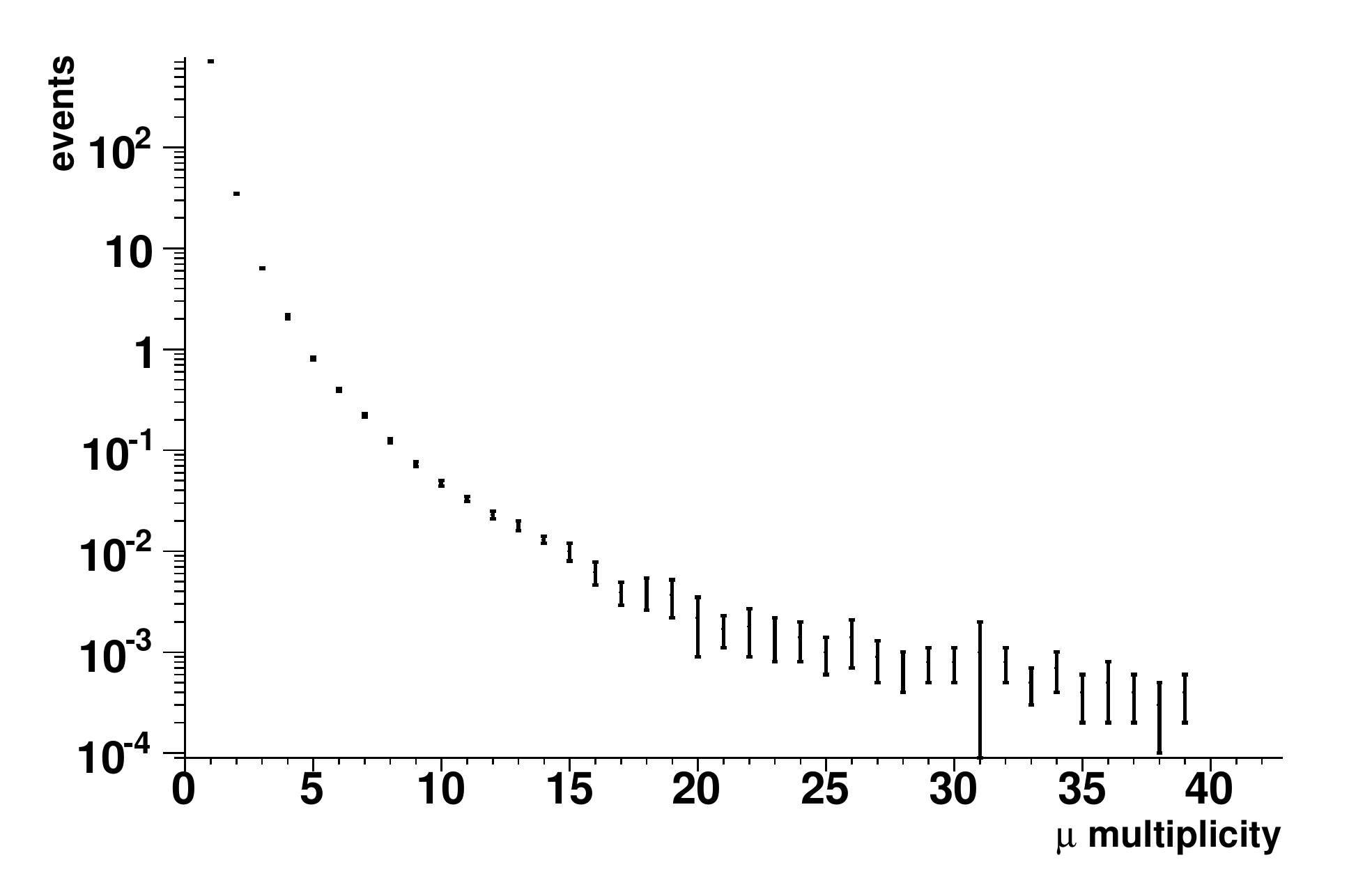}
    \includegraphics[width=.495\textwidth]{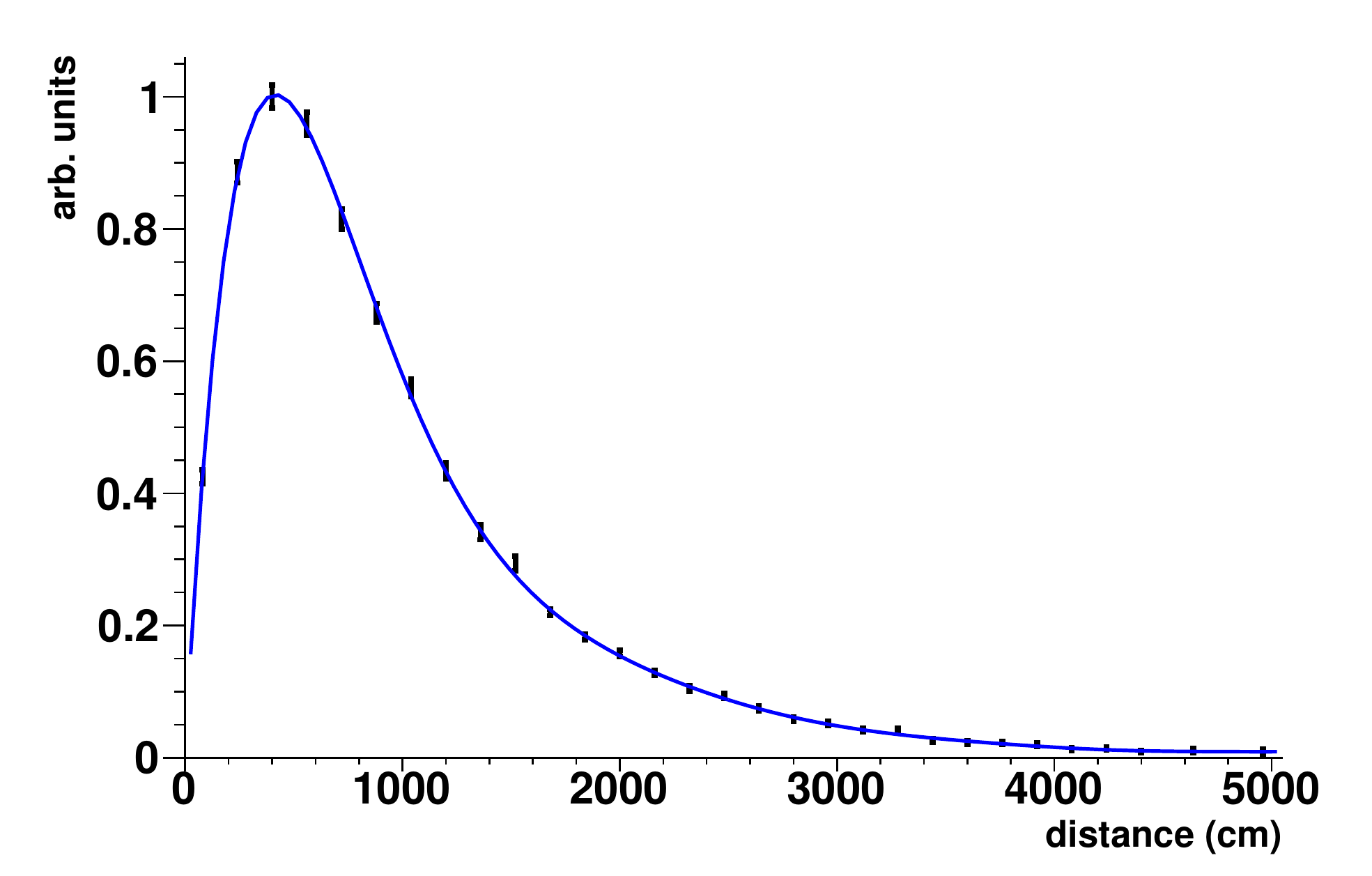}
    \caption{ (left) Multiplicity of muons recorded by the MACRO detector
       for cosmogenic muon events.  \,\, (right) Spatial separation of muons
       obtained for cosmogenic events in the MACRO detector featuring
       two coincident muons (black data points). The blue line shows a
       simple higher order polynomial fit used to implement sampling
       of the distribution. }
   \label{fig:bundle}
   \end{figure}

 Muon bundles were investigated at LNGS by the LVD and MACRO experiments.
 The experimental results used here were taken from MACRO~\cite{bundles}.
 Figure~\ref{fig:bundle} shows the measured muon multiplicity (left) and
 the spatial separation between muons for double muon events (right).  A
 simulation to study muon multiplicity used simplified
 sampling of the distribution up to a muon multiplicity of 4.  The
 distance between muons within a bundle was chosen according to the
 distribution measured for double muon events, and all muons within a
 bundle are given the same direction. 

 The effect of multi-muon events for the Borexino detector geometry
 was evaluated assuming the measured multi-muon event rate from MACRO of
 approximately 6\%.   Close to 1.5\% of the muon events in Borexino
 feature more than one muon.   In addition, about 12\% of the single muons
 crossing the Borexino inner detector belong to multi-muon events.

\subsubsection{Event generation}

 To generate the muon flux within the experimental hall,
 a combination of azimuthal 
 and zenith angles is selected according to the measured muon angular 
 distribution.   A map of the LNGS (Gran Sasso mountain) overburden,
 which was prepared 
 by the MACRO collaboration \cite{PhysRevD.52.3793,max}, is used to
 translate the muon direction into the respective slant depth $h$. 
 Next, the muon event type is set to either a single muon or muon bundle.
 In the case of muon bundles, a multiplicity of up to 4 muons is
 sampled from the measured multiplicity spectrum.   The probability
 for muon events with larger multiplicity is less than 0.2\% and these
 events are treated as muon bundles of multiplicity 4 to simplify the
 calculation.

 Finally, the muon kinetic energy as a function of slant depth $h$ and
 muon event type is selected by sampling from the parameterized
 single or double muon event energy spectra.  The latter is used for
 muon bundle events of all multiplicities since no experimental
 information is available for events with higher muon multiplicities.

 An important difference between positive and negative muons arises
 from negative muon capture on nuclei.  This effect can produce
 energetic neutrons, in some cases over 100 MeV. However, at
 the LNGS depth the fraction of stopping muons is less than 1\%.
 A constant charge ratio of N$_{\mu_{+}/\mu_{-}}=1.38$ was selected
 to simplify the simulation.  This value is consistent with the
 weighted average of the reported measurements for single and
 multi-muon events by OPERA~\cite{opera}: R$_{single}=1.395\pm0.025$
 and R$_{multi}=1.23\pm0.1$.

 \subsection{Muon-induced secondaries}

 \subsubsection{Geometry}

\begin{figure}[hbt]
   \centering
  \includegraphics[width=.495\textwidth]{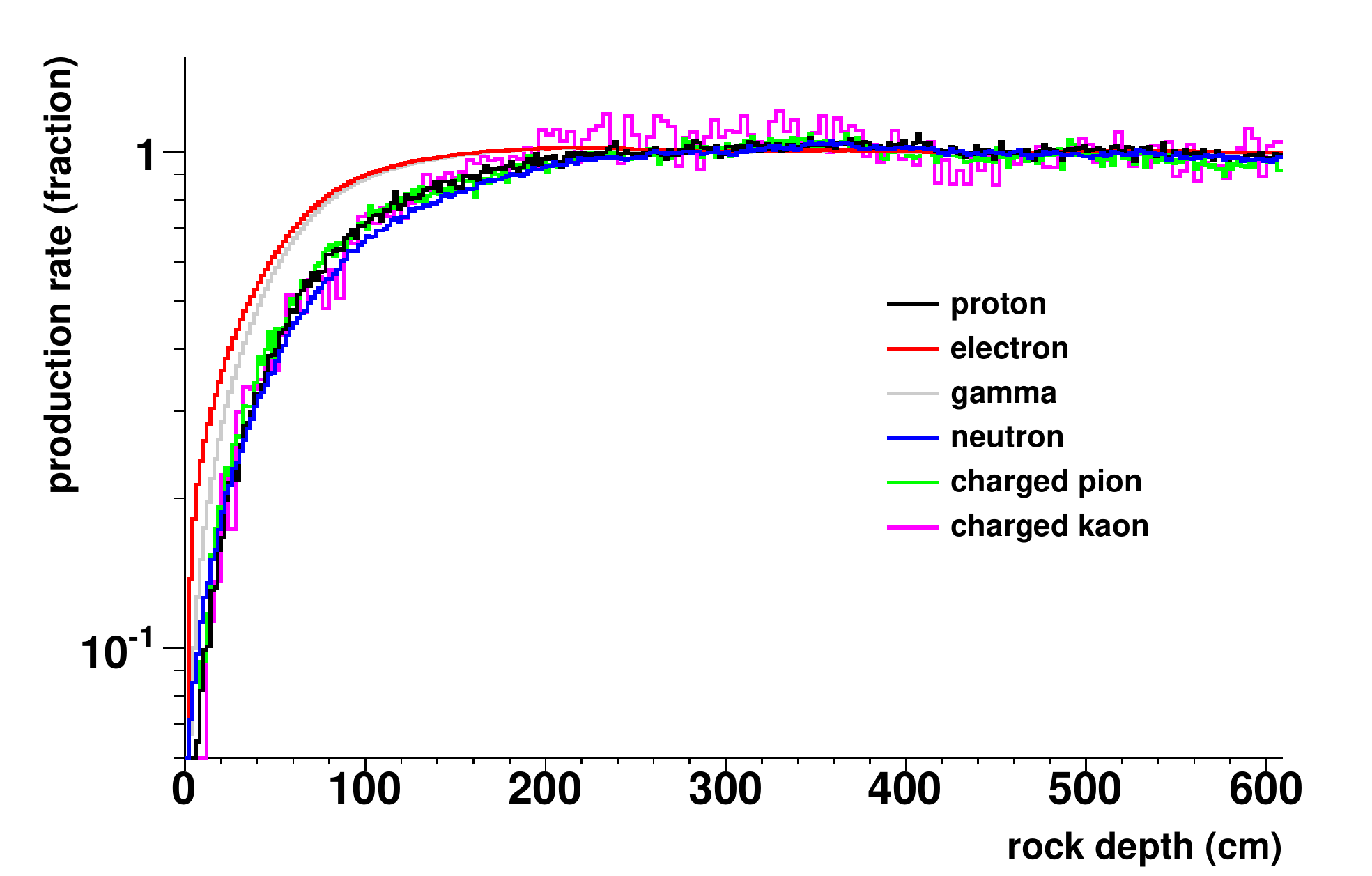}
  \includegraphics[width=.495\textwidth]{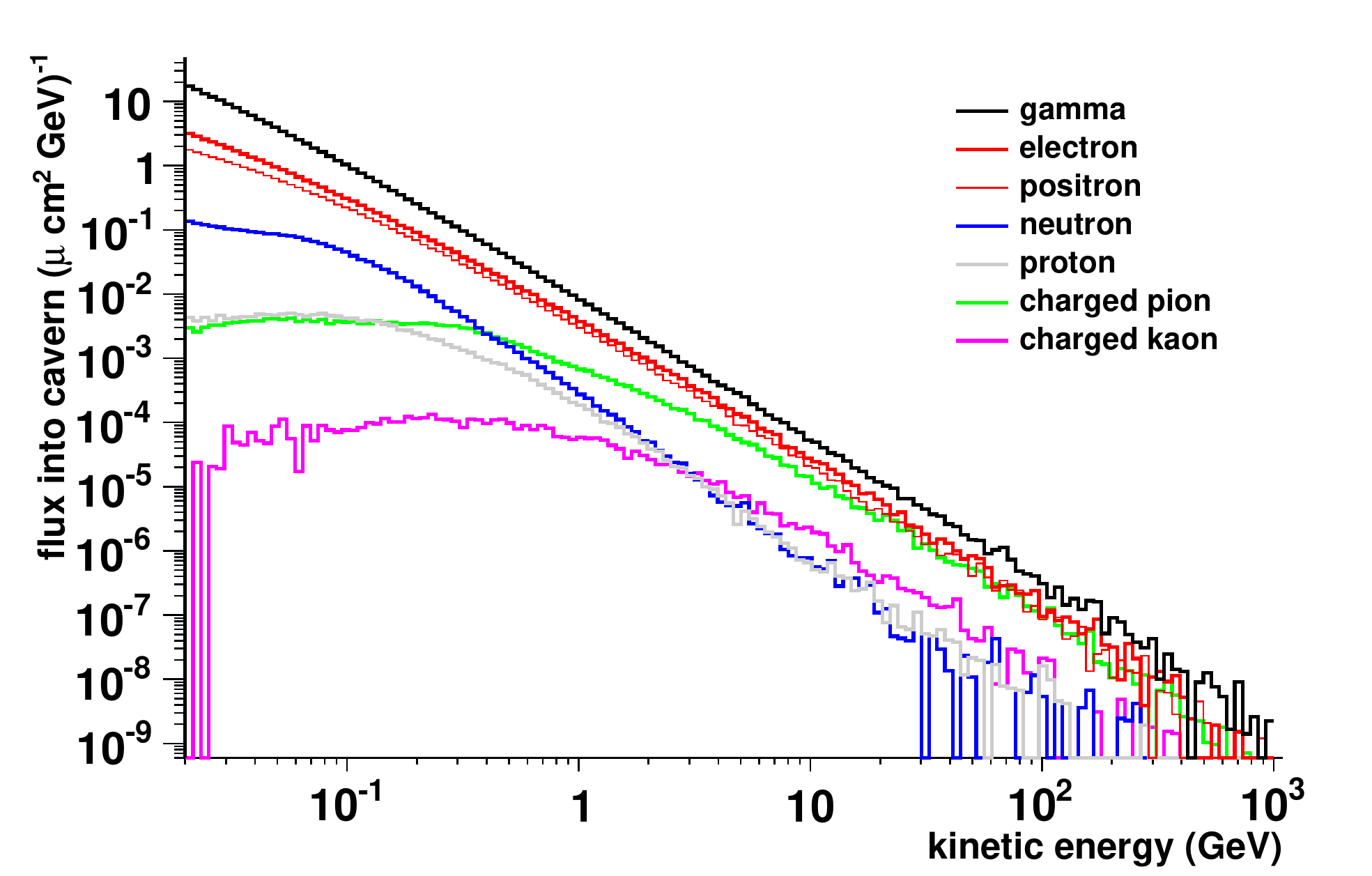}
    \caption{
         (left) Particle production rates by 280 GeV muons in Gran Sasso rock
          as function of distance travelled.  The rates are normalized per
           particle species to the maximum production rate. \,\,
         (right) Predicted integral particle flux into Hall C at LNGS per
           cosmogenic muon event given as a function of particle kinetic energy.
           }
   \label{fig:particles}
   \end{figure}

 The cosmogenic radiation field at deep underground sites is composed
 of muons and muon-induced secondaries.   Incident muons are allowed
 to develop particle showers as they pass through a 700~cm thick
 layer of Gran Sasso rock \cite{PhysRevD.52.3793} surrounding all
 sides of Hall C.   The amount of rock to fully develop the shower
 was determined by simulation.  Particle production rates for muons
 of 280 GeV kinetic energy in Gran Sasso rock are reported in
 Figure~\ref{fig:particles} (left).  Constant particle production rates,
 indicating full shower development, are reached for a rock thickness
 of 300-400 cm.  However for computational
 purposes, this thickness was divided into three sections to permit
 simulation of electromagnetic processes with increasing detail as
 the shower approached the cavern.

 Muon events are randomly placed on a sufficiently   large   plane located
 above the hall geometry in order
 to fully illuminate the hall containing the detector.  Other muon tracks
 outside this envelope are rejected.  The basic assumption of this
 procedure is to implement an initial distribution
 of muons at every point within the rock with appropriate
 direction and energy.

 \subsubsection{Propagation through rock}

 The cosmogenic radiation which is present at the cavern walls is
 approximated in two separate steps to reduce computation time.  
 In the first step, the muon radiation field as described in the
 previous section, is reproduced on a rock layer surrounding Hall C, and
 allowed to propagate without interactions through the rock.  These muons
 are recorded if they
 entered the hall.  For high energy, the muon trajectory is
 approximately unchanged by interactions, so that these recorded muons
 were chosen as the muon sample for the simulation.  They were
 then propagated a second time with all physics processes
 enabled in order to obtain a description of the cosmogenic radiation
 field, including all muon-induced secondaries.  The secondaries
 were recorded along with the muons as part of the incident flux.
 The sampled muon kinetic energy was adjusted for the average muon energy
 loss in the rock layer.

 \subsubsection{Particle components of the muon-induced radiation field}

 As previously indicated, high energy muons produce many particle
 types in addition to neutrons. These particles, including the primary
 muon, can continue to produce backgrounds as they interact with the
 detector and its surroundings.  Figure~\ref{fig:particles} (right)
 shows the kinetic energy spectra for the most frequently produced
 secondaries.  Aside from muons, photo-production
 can also contribute to backgrounds, including neutron backgrounds,
 due to the large photon flux even
 though electromagnetic cross sections are small.
 In order to assess cosmogenic backgrounds, the full muon-induced radiation
 field needs to be considered rather than, for example, simulating only
 cosmogenic neutrons.  The character of the background problem may be
 significantly changed considering additional coincident charged secondaries.

 \subsection{Connection of simulation to measured muon event flux}

 The muon flux per simulated event, $\Phi_{sim}$ as determined above,
 is then allowed to interact with the detector and shielding.
 An empty spherical volume was inserted inside the cavern, and exposed
 to the muon radiation field in order to determine the flux incident
 on the Borexino detector.  The muon fluence estimated by the track
 length-density inside the sphere was obtained by a standard FLUKA
 scoring option.   Choosing a spherical volume correctly accounts for
 the angular distribution of
 the muons.  Moreover, the fluence  rate  through a sphere is a direct
 estimator of the flux through the cross-sectional area of the sphere.

 The length of the time-period considered in the simulation (lifetime),
 is given by the number of simulated muon events compared to
 the ratio of the simulated to measured total muon flux,
 $\Phi_{exp}$, from Table~\ref{mu_flux}.

 $$ T [s] = N_{events} \cdot { \Phi_{sim} [events^{-1} cm^{-2}] \over{
 \Phi_{exp} [s^{-1} cm^{-2}] }} $$


 \section{Validation}

 \subsection{The Borexino experiment}

 The most recent, and also the most precise experimental data on cosmogenic
 neutrons deep underground, are available from the Borexino
 experiment~\cite{cosmogenic_bx}.  The reduced systematic uncertainties
 of the reported results are primarily a consequence of the very
 large detector with its un-segmented and shielded spherical liquid
 scintillator target.  In addition, Borexino has a comparatively
 short recovery time from the large prompt muon signal.  This allows
 less extrapolation and less accidental background between the muon and
 capture signals.
 The experimental results
 and a comparison with Monte Carlo predictions are presented
 in~\cite{cosmogenic_bx}.  These simulations were based on FLUKA
 using the muon radiation field at LNGS as developed
 in the previous section.


  \begin{figure}[tbh]
   \centering
   \includegraphics[width=.6\textwidth]{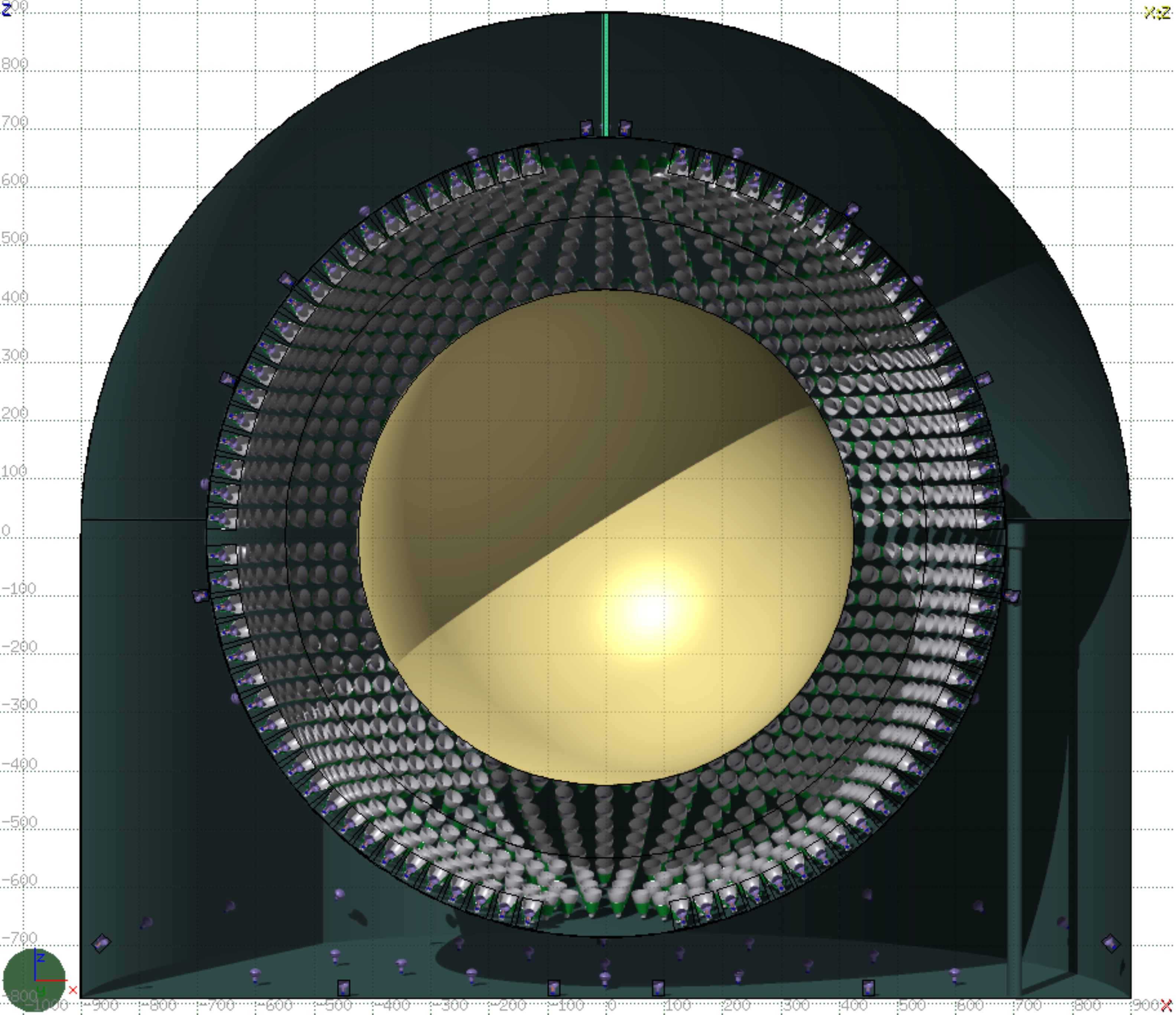}
   \caption{ A view of the Borexino detector geometry implemented in FLUKA.
             The sensitive volume of the Borexino detector, the Inner
             Vessel, is shown by the golden sphere.  It is centered inside
             two (transparent) buffer regions which are contained inside
             a stainless steel sphere which supports the optical modules.
             This inner detector is placed inside a domed, cylindrical
             water tank which functions as a \v{C}erenkov detector to
             identify and help track cosmogenic muons.
            }
  \label{fig:bx}
  \end{figure}

 A cross-sectional view of the FLUKA geometry implemented for the Borexino
 experiment is given in Figure~\ref{fig:bx}.  The central sensitive
 liquid scintillator region of radius 425~cm has a mass of
 approximately 278 tons and is contained in a spherical, nylon inner
 vessel.  The detector is located at the center of a 685~cm radius
 stainless steel sphere filled with quenched liquid scintillator
 as a passive shield.  The stainless steel sphere in turn is placed
 inside a 900~cm radius cylindrical water tank providing additional shielding
 as a water \v{C}erenkov muon veto detector.  The
 detector components are carefully modeled in the simulation.
 The liquid
 scintillator is Pseudocumene with the addition of a 1.5 g/l
 fraction of wavelength shifter 2,5-diphenyl-oxazole (C$_{15}$H$_{11}$NO).
 The FLUKA simulation also implemented the photo-sensors of the
 inner detector, but no attempt was made to include optical transport, or
 to address the creation of detector signals                                      
               at this point because of computational time considerations.        
 However, the production of cosmogenic isotopes as well as neutron capture
 reactions on hydrogen inside the Inner Vessel were recorded.

 \subsection{Benchmark results}

 \subsubsection{Muon-induced neutron yield}

 The concept of neutron yield was given in~\cite{zat65} as ``the number of
 neutrons produced by one fast muon per g cm$^{-2}$'', and predicted by theory
 to increase with the mean muon energy approximately as
 $\overline{E_{\mu}}^{\,0.7}$.  However as defined, this quantity is not
 accessible to experiment.  Underground experiments record the
 capture of thermalized neutrons in liquid scintillator detectors.  In part,
 this is necessary because of the large initial energy deposited by the parent
 muon(s) so that only delayed neutron capture signals can be identified.
 Borexino reports the cosmogenic muon-induced thermal neutron capture yield in
 liquid scintillator, ignoring fast neutron capture reactions on carbon
 which are predicted to only account for approximately 1\% of neutron captures.
 Further, a fraction of the created energetic
 neutrons re-interact, and not all nuclear reactions will have neutrons
 in the final state.  The predicted discrepancy between the total number
 of neutrons created during cosmogenic events as opposed to the number
 of thermal neutron captures, is somewhat less then 10\% for a liquid
 scintillator target.  This discrepancy can only be addressed through
 modeling and is distinct from the frequently mentioned issue of ``double
 counting neutrons'', which in simulation work only refers to (n,xn)
 type processes, with xn indicating $\ge$2 final state neutrons.
 Early published experimental results did not address this difference.  In
 addition, the early experiments tried to assess the neutron production
 yield in standard rock rather than for liquid scintillator.

 The reported experimental and FLUKA predicted neutron capture yields for
 Borexino~\cite{cosmogenic_bx} are:

 \begin{table}[hbt]
   \begin{center} \begin{tabular}{lc}
                  &         Yield  $ [\times 10^{-4} (muon\,\,  g/cm^{2})^{-1}] $  \\
     \hline
     Borexino           &  $ 3.10 \pm 0.11 $ \\
     FLUKA prediction   &  $ 2.46 \pm 0.12 $ \\
   \end{tabular} \end{center}
  \vskip -3mm
  \caption{Cosmogenic muon-induced neutron capture yield. Only
  statistical errors are given for the simulation.}
 \end{table}

\begin{figure}[thb]
   \centering
   \includegraphics[width=.495\textwidth]{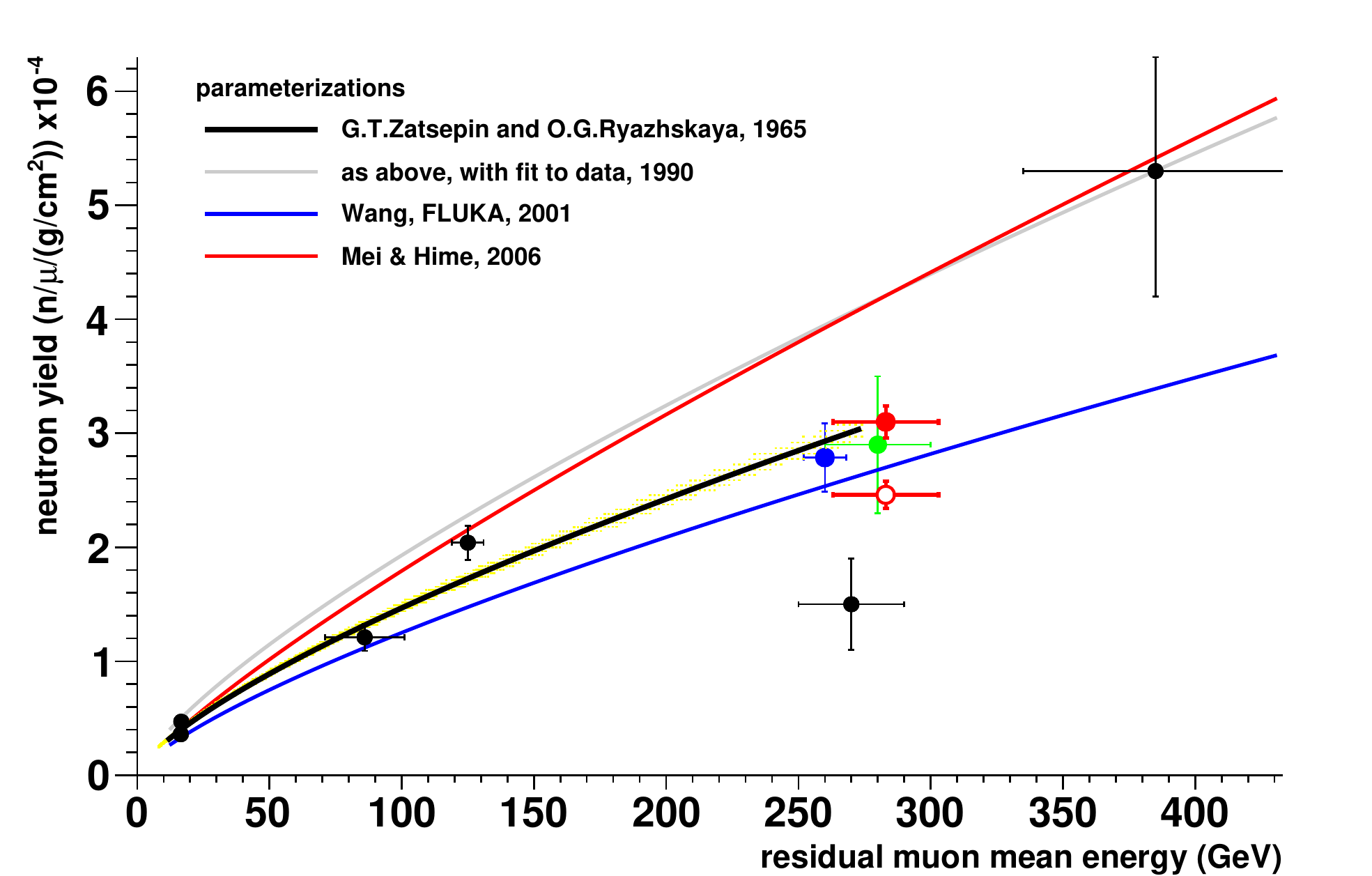}
   \includegraphics[width=.495\textwidth]{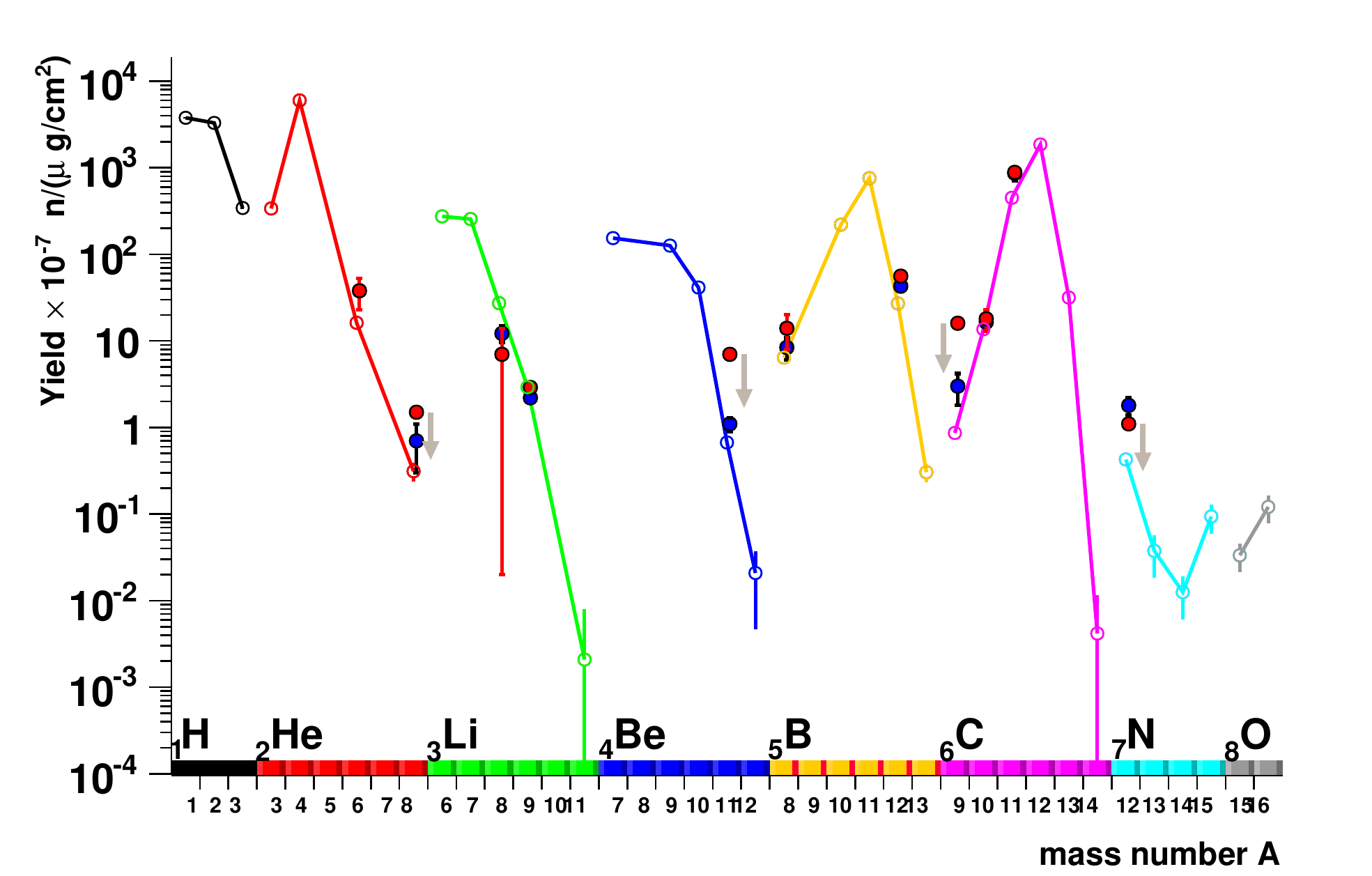}
   \caption{ (left) Neutron capture yield as a function of mean muon energy with
               various representations of the experimental data.
             (right) Muon-induced cosmogenic isotope production yields
             in liquid scintillator at LNGS energies.  Experimental
             results from Borexino (solid red symbols) and KamLAND
             (solid blue symbols) for some unstable isotopes are
             compared to FLUKA predictions.  The Gray arrows indicated reported
                upper limits.
            }
   \label{fig:iso}
   \end{figure}

 \hskip -\parindent
 The measured and the simulated neutron capture yields are shown by the
 solid and open red symbols, respectively, as a function 
 of mean muon energy on the left in Figure~\ref{fig:iso}.  They are compared
 to the available experimental data
 on neutron production in liquid scintillator.  The data are the originally
 published values of neutron yield and mean muon energy~\footnote{The result
 reported by the Palo Verde experiment~\cite{boehm} did not provide a mean
 muon energy and the value reported in~\cite{wang} is adopted here.}.  The
 solid black symbols correspond to the historical measurements at increasing
 mean muon energy in GeV of 16.5~\cite{boehm}, 16.7 and 86~\cite{bezrukov0},
 125~\cite{enikeev}, 270~\cite{lvdAgli} and 385~\cite{lsdAgli}.
 More recent values are presented by the solid blue and green symbols from the
 KamLAND~\cite{kamland} and LVD~\cite{bib:persiani} experiments, at mean muon
 energy of 260 and 280 GeV respectively.

 The original theoretical prediction for the neutron yield in standard rock at
 depth is shown by the thick black line~\footnote{ Specifically, the calculations       
 were carried out for an aluminum target.   
 However, predictions for standard rock should be reduced on the order 10\%. }.         
                                          The prediction predates experimental
 information.  It was stated that: ``such calculations give the dependence on
 $\overline{E_{\mu}}$ to good accuracy, but absolute values are accurate only
 in the order of magnitude''~\cite{zat65}.
 Later fits of the predicted power law to the early data resulted in
 parameterizations of the neutron yield, as
 for example, reported in \cite{lsd1990} and depicted by the thin gray curve.
 The blue curve shown is a parameterization which was derived
 from an early FLUKA simulation~\cite{wang}.

 Finally, the red curve is added which
 compares the predicted neutron yield as suggested
 by~\cite{hime}.  This work is frequently quoted in context of simulations
 for cosmogenic muon-induced neutrons deep underground where the FLUKA
 predictions
 were modified by arbitrarily post-scaling the muon-induced neutron production,
 as a function of mean muon energy, in order to match the historical
 data.  More recent data have revised the experimental yield downward
 comparable to the unmodified FLUKA prediction.

 Extra care is required to interpret the early experimental data.
 The assigned experimental uncertainties seem optimistic when taking into
 account the less refined level of experimental information available at the
 time, and the lack of detailed Monte Carlo simulations. This is important
 since the early detectors were smaller and not well shielded and/or constructed
 with large amounts of mixed passive target materials in addition to liquid
 scintillator.  Similar complications apply to the assigned mean muon energies.
 A reduced neutron yield is supported by the original theoretical prediction
 together with new quality measurements at mean muon energies of approximately
 260~GeV and 283~GeV ~\cite{cosmogenic_bx}. Note that FLUKA predictions
 and experimental data for the muon-induced neutron yield in liquid
 scintillator are now in fair agreement.
 These results are specific to light target materials and do not apply to
 heavy targets made of iron or lead.

 \subsubsection{$^{11}$C isotope production}

 The creation of $^{11}$C by cosmogenic muons in liquid scintillator was
 studied extensively by Borexino~\cite{pep,cosmogenic_bx} and the most recent
 experimental result is reported in Table~\ref{c11yield}.  Approximately 30\%
 of the cosmogenic neutrons measured in Borexino arise as final state neutrons
 in interactions leading to the creation of $^{11}$C nuclides.
 \begin{table}[thb]
   \begin{center} \begin{tabular}{lc}
                  &         Yield  $ [\times 10^{-4} (muon\,\, g/cm^{2})^{-1}] $  \\
     \hline
     Borexino                                        &  $ 0.886 \pm 0.115 $ \\
     FLUKA prediction                                &  $ 0.467 \pm 0.023 $ \\
   \end{tabular} \end{center}
  \vskip -3mm
  \caption{Cosmogenic muon-induced production yield of $^{11}$C. Only
  statistical errors are given for the simulation.}
  \label{c11yield}
 \end{table}
 Also listed in Table~\ref{c11yield} is the FLUKA predicted $^{11}$C yield.         
                                                              Only half as
 many $^{11}$C nuclei are predicted as compared to the Borexino measurement.
 One reason for the low $^{11}$C production
 rate in FLUKA is addressed by improvements to the
 Fermi break-up model~\cite{fluka10} and these will be available with the next
 FLUKA release.
 In FLUKA, nuclear de-excitation is performed according to a Fermi break-up
 model for light nuclei with mass number A$\le$16.
 The updated model implements additional conservation laws as well as
 constraints on available final state configurations and symmetries.  In the
 case of $^{11}C$, the $ \gamma\,+\,^{12}C $ reaction in the current FLUKA
 model strongly favors break-up of $^{12}C$ into 3\,$\alpha$ over neutron
 emission via photo-production.  However, the parity and spin of the
 3 $\alpha$ breakup is:
    $$ \gamma \, (1^{-}) + ^{12}C \, (0^{+}) \,
               \rightarrow \,3 \,\alpha (\, 0^{+}). $$
 Thus, the  breakup into 3\,$\alpha$ particles with L=0 is forbidden
 even though this breakup is energetically favored.  As a result, the reaction \,
 $^{12}$C($\gamma,n$)$^{11}$C \, is under-predicted by the current and previous
 FLUKA versions.  An increase in the photoproduction of $^{11}$C by a
 factor of 2 to 3 is expected with the new model, see
 Figure 4 in~\cite{fluka10}.
 The production of $^{11}$C was studied with a beta version of FLUKA
 and the more than twofold increase of the $^{11}$C new photoproduction rate was
 confirmed.  Different reactions triggered by cosmogenic muon-induced
 secondaries also produce $^{11}$C nuclei in liquid scintillator.  The
 fraction of inclusive photoproduction contributes with 64\% and the
 improved prediction for the $^{11}$C production yield
 is (0.70$\,\pm\,$0.02)$\times 10^{-4} [(muon\,\, g/cm^{2})^{-1}]$.  Still this
 result is about 30\% low with respect to experiment.  The
 predicted fraction of $^{11}$C production without a final state neutron
 is close to 10\%.

 Taking the improved $^{11}$C production model into account, one
 could expect
 the predicted neutron capture yield to increase to approximately
 2.7$\times 10^{-4} [(muon\,\, g/cm^{2})^{-1}]$, which is within 15\% of the
 reported experimental value.  The FLUKA predictions for $^{11}$C production
 in liquid scintillator can be compared to previous
 estimates~\cite{galbiati} giving the rate of inclusive
 photoproduction at 60\% and the fraction of $^{11}$C production without final
 state neutrons at 5\%.
 The missing, underpredicted neutrons due to the
 low $^{11}$C  yield, are expected to have a ``softer'' spectrum and
 are  more easily removed by shielding.

\subsubsection{Cosmogenic isotope production}

 The production yield for a list of unstable isotopes resulting from cosmogenic
 muon interactions in Borexino together with FLUKA predictions were reported
 in~\cite{cosmogenic_bx}.  A similar study at comparable depth is available
 from the KamLAND experiment~\cite{kamland}.  In Figure~\ref{fig:iso} on the
 right the experimental results are shown by the solid red and blue symbols for
 Borexino and KamLAND, respectively.  They are compared to FLUKA predictions
 for the cosmogenic isotope production in liquid scintillator.  Only an
 upper limit is reported by Borexino for some yields, and these are indicated in
 the graph by gray arrows.  Good agreement between data and simulation is
 found considering the large variation in production yield spanning many
 orders of magnitude and the complexity involved in the simulation.  As
 discussed in the previous section, the agreement for the $^{11}$C production
 will improve with the next version of the FLUKA code, and predictions
 for other light isotopes may also be affected by these changes.

 A recent FLUKA study for the Kamiokande water \v{C}erenkov detector
 also reports good agreement for cosmogenic isotope production in a
 water target \cite{shirley}.  Liquid scintillator and water are
 two of the shielding media used for typical underground experiments
 and will be employed in this work.


 \subsubsection{Distance between neutron capture location and muon track}

 No information about the prompt neutron yield can
 be directly obtained from scintillation detectors due to the
 large initial signal caused by a penetrating muon(s) through the detector.
 However, the spatial reconstruction of the muon track and
 the delayed neutron capture locations in Borexino facilitate a study of the
 transverse distance neutrons can travel perpendicular to the parent
 muon track.  In Borexino, only a single
 muon track is reconstructed per cosmogenic event since muon-bundles
 are not identified.

\begin{figure}[htb]
   \centering
   \hskip -2.5mm
   \includegraphics[width=.495\textwidth]{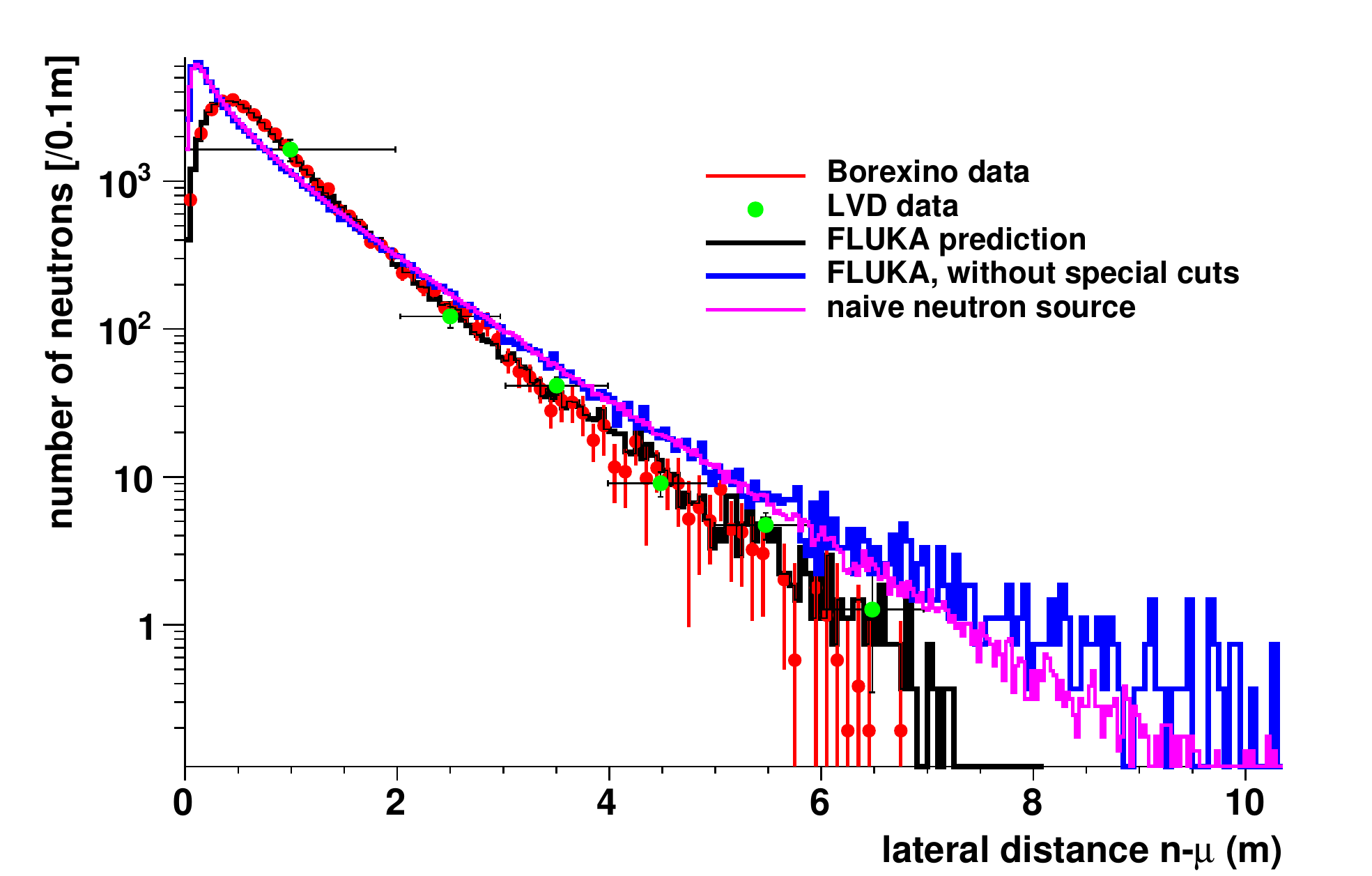}
   \includegraphics[width=.495\textwidth]{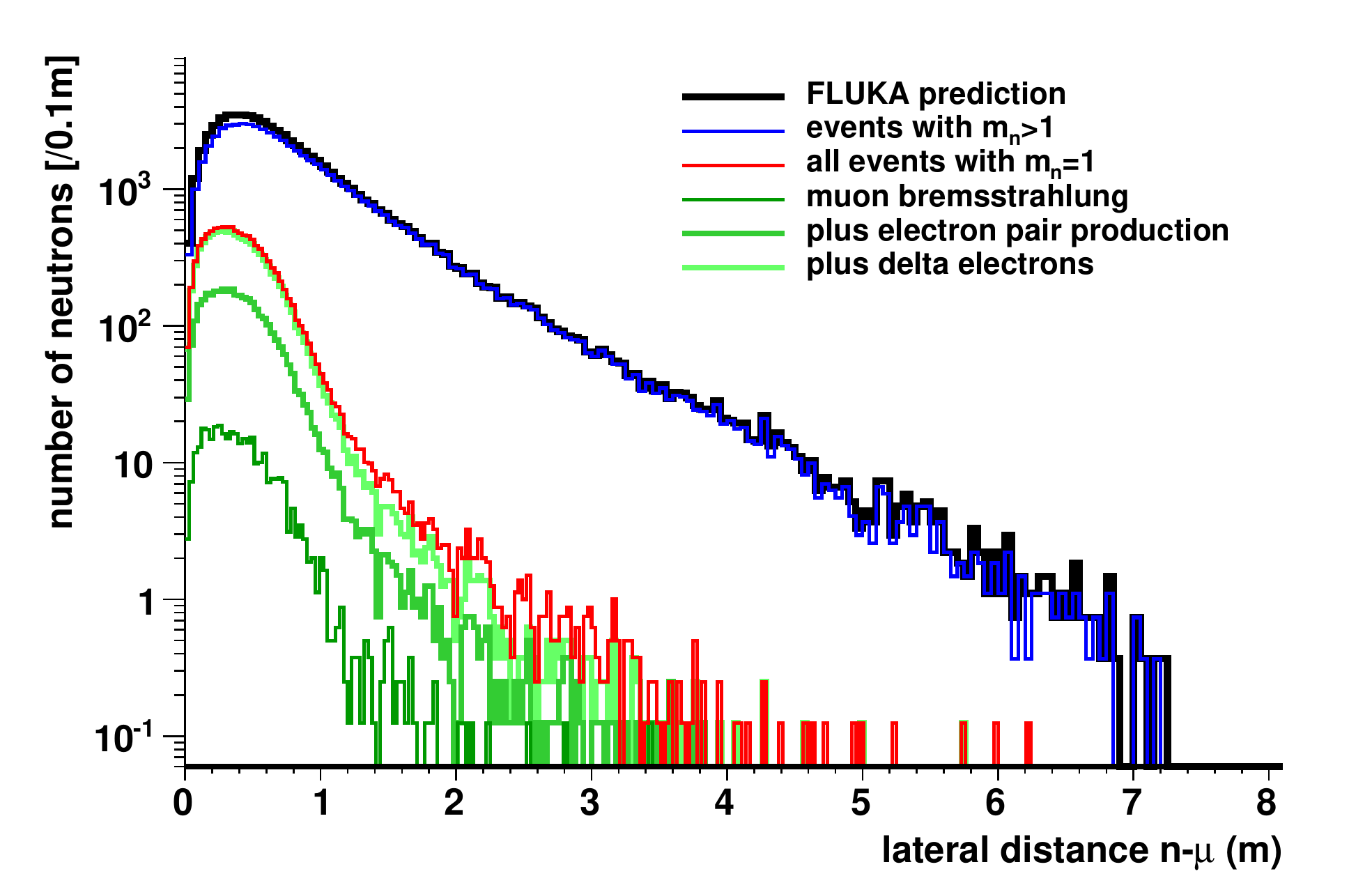}
   \caption{ (left) Borexino experimental result, red symbols, and FLUKA
             predictions, black histogram, as reported in \cite{cosmogenic_bx}
              are compared to a measurement from LVD~\cite{lvdlat}, solid
               green symbols.  The blue histogram gives the predicted lateral
                distribution before applying reconstruction uncertainties
                 and additional radial cuts.  This histogram is compared
                  to the lateral distribution found from FLUKA for a simple
                   neutron simulation (purple line), see text.
          (right) The predicted lateral neutron capture distribution, black
            histogram, is composed of neutrons from muon events producing many
             or single neutrons shown by the blue and red histograms
              respectively.  For the latter, the cumulative fractions of events
               initiated by muon bremsstrahlung (dark green) electron pair
                production (green) and delta ray production (light green)
                 are indicated.  }
  \label{fig:lateral}
  \end{figure}

 The lateral distribution of neutron captures from the parent muon track is
 shown in Figure~\ref{fig:lateral}.  All the curves are normalized to
 permit comparison with the Borexino results.  Both, the measured (red symbols)
 and the FLUKA predicted (solid black line) distributions from
 \cite{cosmogenic_bx} are reproduced in the graph on the left.
      The muon track and the capture locations were constrained to
 lie within a radius, R$<$400\,cm, with respect to the detector center
 in order to obtain a
 clean data sample.  Reconstruction uncertainties were applied
 {\it a~posteriori} to the FLUKA predictions and these dominate the
 distribution
 at small distances from the parent muon track.  The original distributions
 are further compared to results from the LVD experiment~\cite{lvdlat}
 indicated by the solid green symbols.  The shape of the experimental
 distributions is reproduced by the simulation.

 The predicted lateral distance distribution without
 reconstruction uncertainties and no radial cuts is shown by the blue
 histogram in the same graph.  Here, events
 with single muons crossing the Borexino Inner Detector were selected.
 The general shape of the distribution can readily be reproduced by
 assuming an isotropic neutron production around the muon track with
 an effective neutron diffusion length of approximately 90~cm. Note   that  
 only transverse information is accessible to experiment.
 This diffusion length is a very weak function of the kinetic energy for
 neutron energies above 200 MeV.  Indeed one finds the purple histogram shown
 in Figure~\ref{fig:lateral}, when simulating a neutron beam with a
 flat kinetic energy spectrum (E$_{kin}<$\,350\,MeV) into liquid scintillator,
 and recording the
 lateral distance of thermal neutron capture locations with respect to
 the original neutron direction.  Only at large distances does one see a small
 deviation from the predicted muon-induced neutron capture distance
 spectrum.  However, this difference can be attributed to muon bundle
 events in the data since all neutrons are assigned to a single muon crossing
 the Borexino Inner Detector, and these could have been produced by coincident
 muons from bundles crossing the outer detector.

 The predicted lateral distribution for Borexino is repeated on the right
 in Figure~\ref{fig:lateral} by the solid black histogram.  The
 contributions to this spectrum from events with a single neutron
 capture are shown by the red histogram as opposed to neutrons from
 events with higher neutron capture multiplicities given by the blue
 histogram.  Muon-induced neutrons created in events with neutron
 multiplicity one, are predicted to capture substantially closer to the
 parent muon track.

 The initial, discrete muon energy loss process, which eventually
 leads to the neutron production, was identified for these events.
 The cumulative contribution from muon bremsstrahlung
 (dark green), electron pair production (green) and delta ray production
 (light green) are indicated.  Muon-nuclear interactions with only one
 neutron created in the event together with a $<$\,1\% contamination from
 negative muon capture, account for the small remaining difference
 to the red histogram.


 \subsubsection{Neutron multiplicity}
\begin{figure}[htb]
   \centering
   \hskip -2.5mm
   \includegraphics[width=.495\textwidth]{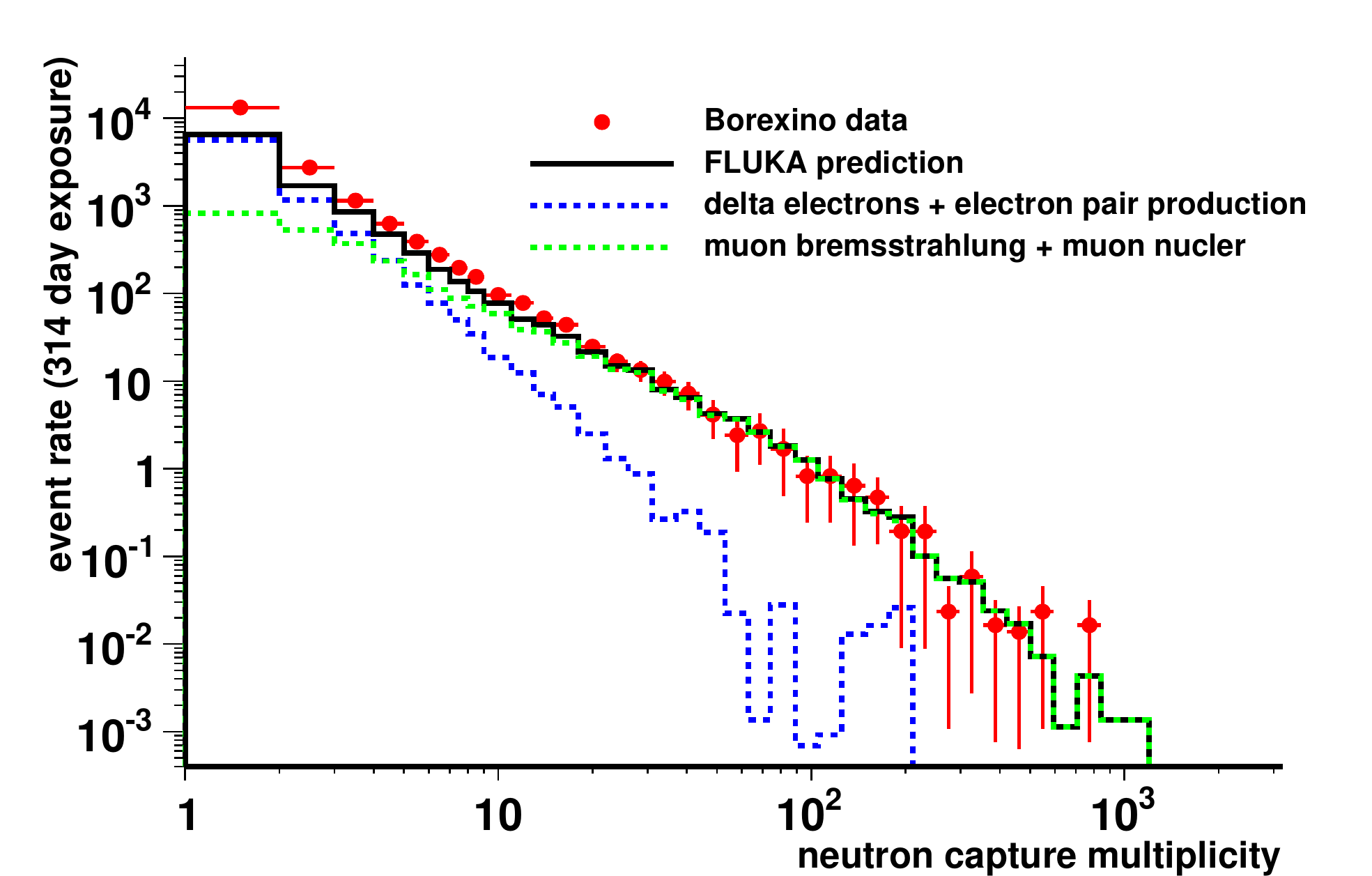}
   \includegraphics[width=.495\textwidth]{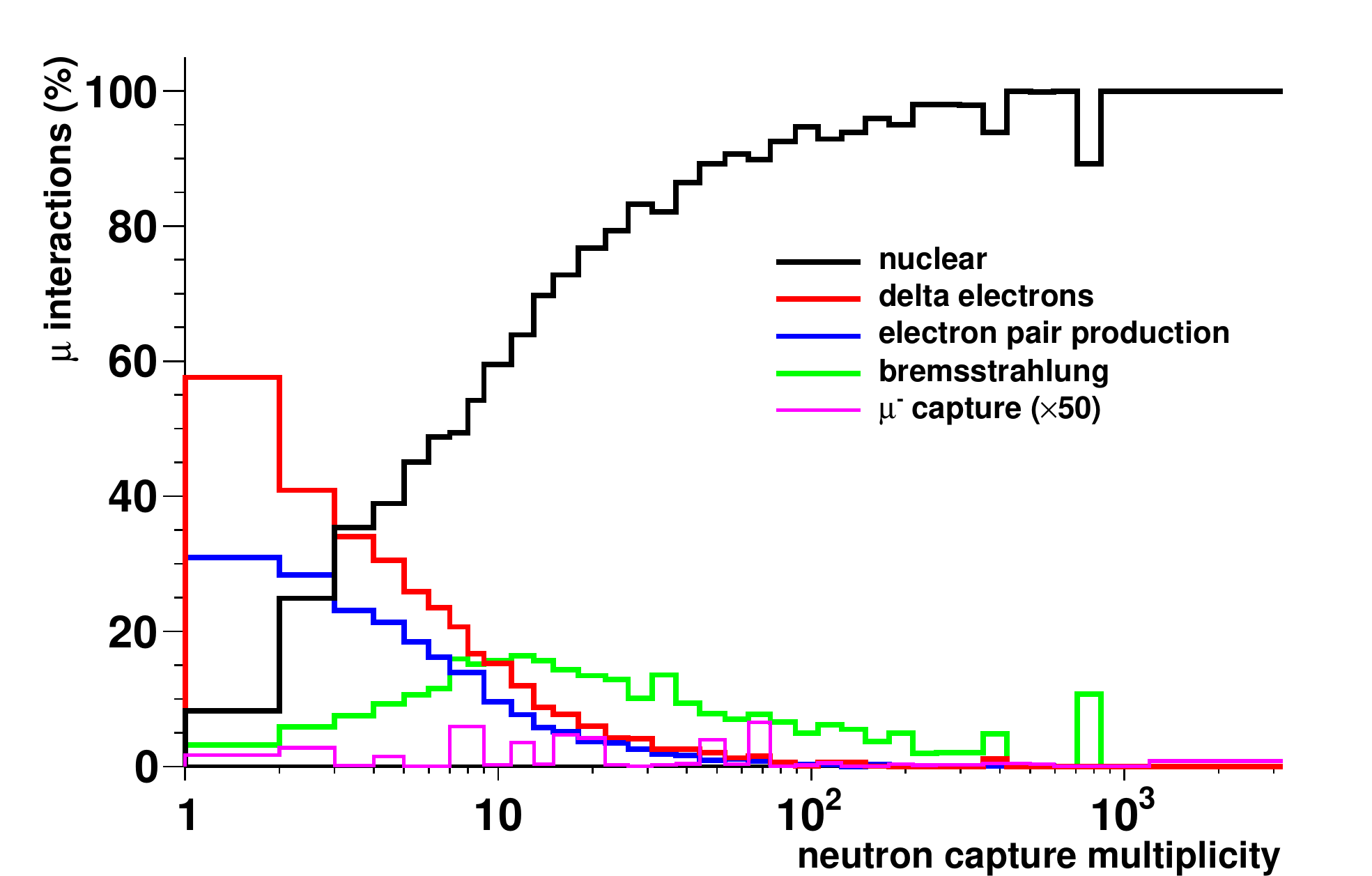}
   \caption{ (left) Absolute comparison of muon-induced cosmogenic thermal
              neutron capture multiplicity as measured by Borexino,
               red symbols, and predicted by FLUKA, black histogram.  For
                the simulated spectrum contributions from muon interactions
                 associated with ``soft'' (dashed blue line) and ``hard''
                  (dashed green line) energy loss processes are shown separately.
             (right) Fractions of different muon interactions which trigger the
              production of neutrons in liquid scintillator as a function of
               neutron capture multiplicity.  The fraction of $\mu^{-}$ capture
                is magnified by a factor of 50 for visibility.
           }
  \label{fig:multiplicity}
  \end{figure}

 In Figure~\ref{fig:multiplicity} the multiplicity of thermal
 neutron captures per muon-induced cosmogenic event is shown.
 Both the Borexino experimental
 result (red symbols) and the FLUKA predicted distributions (black
 histogram) from \cite{cosmogenic_bx} are reproduced in the graph on the
 left.   This comparison is absolute which is different from that of the
 lateral distance distribution.
 The measured distribution is biased
 at large multiplicities due to detector performance with respect
 to energetic and/or muon bundle events.  This effect was best reproduced
 in the simulation by selecting only events with single muon tracks
 crossing the Borexino inner volume.
 Good agreement in the shape of the
 distribution is found except at very low neutron multiplicity events.

 ``Hard'' energy losses by energetic muons, which in turn can produce
 nuclear showers, are more likely initiated by muon bremsstrahlung and
 muon-nuclear interactions.  The contribution to  the  neutron multiplicity
 spectrum attributed to both muon bremsstrahlung and muon-nuclear
 interactions is shown by the dashed green line.  ``Soft''
 energy losses on the other hand proceed mainly via electron pair
 and delta electron production (ionization).  Their contribution to the
 neutron multiplicity spectrum is shown by the dashed blue line.
 The fraction of the individual muon interaction types which result in the
 capture of cosmogenic neutrons as a function of neutron capture
 multiplicity are given in the graph on the right in
 Figure~\ref{fig:multiplicity}.  Almost 90\% of the events which feature
 a single neutron capture are triggered by muons after delta electron and
 electron pair production.


 The electromagnetic interaction of muons is well known and this
 has been carefully benchmarked in FLUKA.
 As reported in section \ref{fluka.models}, muon-nuclear interactions in
 FLUKA are simulated via virtual photon interactions.  However, only
 incoherent interactions on single nucleons are simulated at low energy
 while coherent muon interactions, with multiple nucleons are expected
 to contribute.
 These ``soft'' muon-nuclear interactions are not implemented in the
 present FLUKA code, and may account for at least some of the
 underpredicted yield in Figure~\ref{fig:multiplicity}(left).

 \subsubsection{Muon rate producing neutrons }

 The reported rate of cosmogenic muon events resulting in one or more
 thermal neutron captures in Borexino is  67\,$\pm$\,1  per day.
 The corresponding FLUKA simulated rate of   41\,$\pm$\,3  per day is
 quite low compared to experiment.  The difference is limited to muon
 events with very low neutron capture multiplicities as is apparent
 from the spectrum shown on the left in Figure~\ref{fig:multiplicity}.
 According to FLUKA, these events are predominantly triggered by muon
 electron pair and delta electron creation, which generally yield
 neutrons with a ``softer'' energy spectrum.  In addition, neutrons created
 in muon events with low neutron multiplicity, capture substantially
 closer to the parent track.  This was shown in Figure \ref{fig:lateral}
 on the right for events with single neutron captures.
   This discrepancy, when compared to low energy, coherent muon-nuclear
   interactions, is still under investigation.
 Hence, even though the FLUKA predicted neutron event rate is
 approximately 30\% low, the class of affected neutrons is less
 problematic in view of cosmogenic background suppression.
 Furthermore, because only those events with very low neutron
 multiplicities are affected, the impact on the neutron yield is
 less pronounced.


 \section{Cosmogenic background predictions for \darkside}

 The \darkside experiment is installed inside the Counting Test Facility
 (CTF) which is located in Hall C at LNGS adjacent to the Borexino detector.
 Because of the proximity of these two experiments, the same muon induced
 cosmogenic radiation field which was prepared for Borexino can also be used
 to simulate the cosmogenic background for \Darkside.
 This presents the unique opportunity to validate the
 simulation ansatz, to normalize the predictions to the {\it in-situ} measured
 muon flux, and to study systematic uncertainties of the simulation procedure
 for a ton-sized dark matter experiment.

 \subsection{Detector geometry}

\begin{figure}[bht]
    \centering

    \includegraphics[width=.495\textwidth]{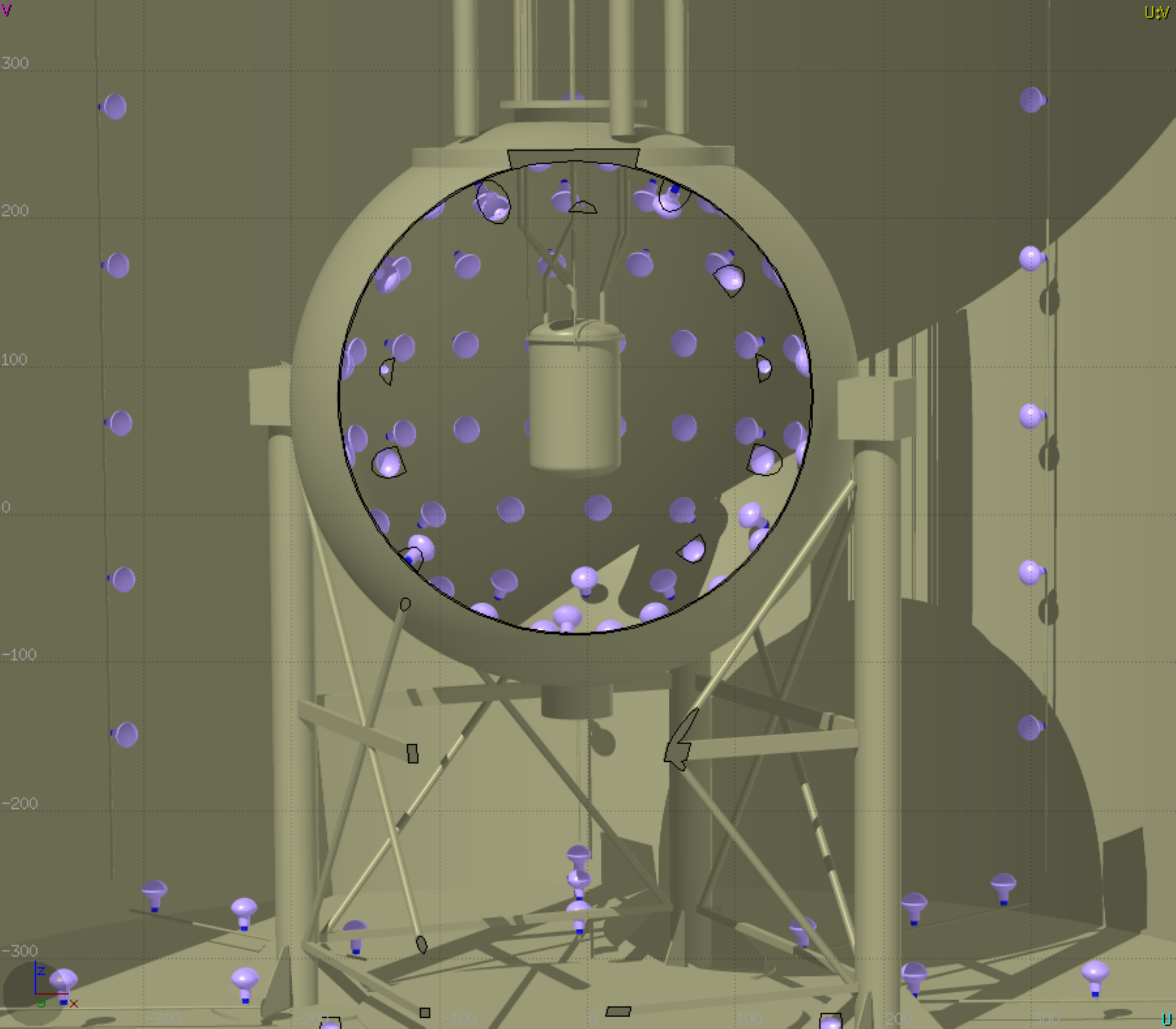}
    \includegraphics[width=.495\textwidth]{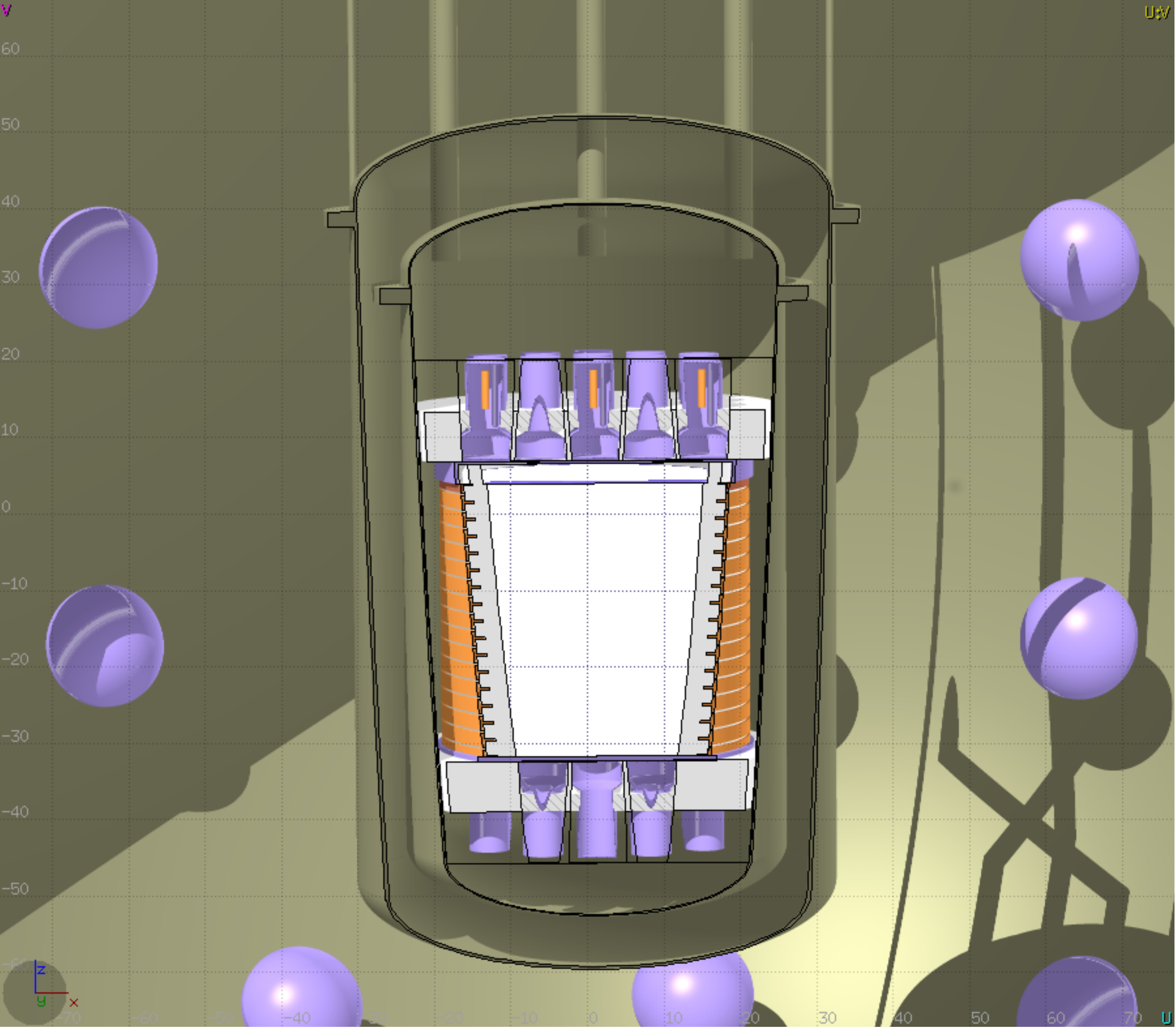}
    \vskip 1mm
    \includegraphics[width=.495\textwidth]{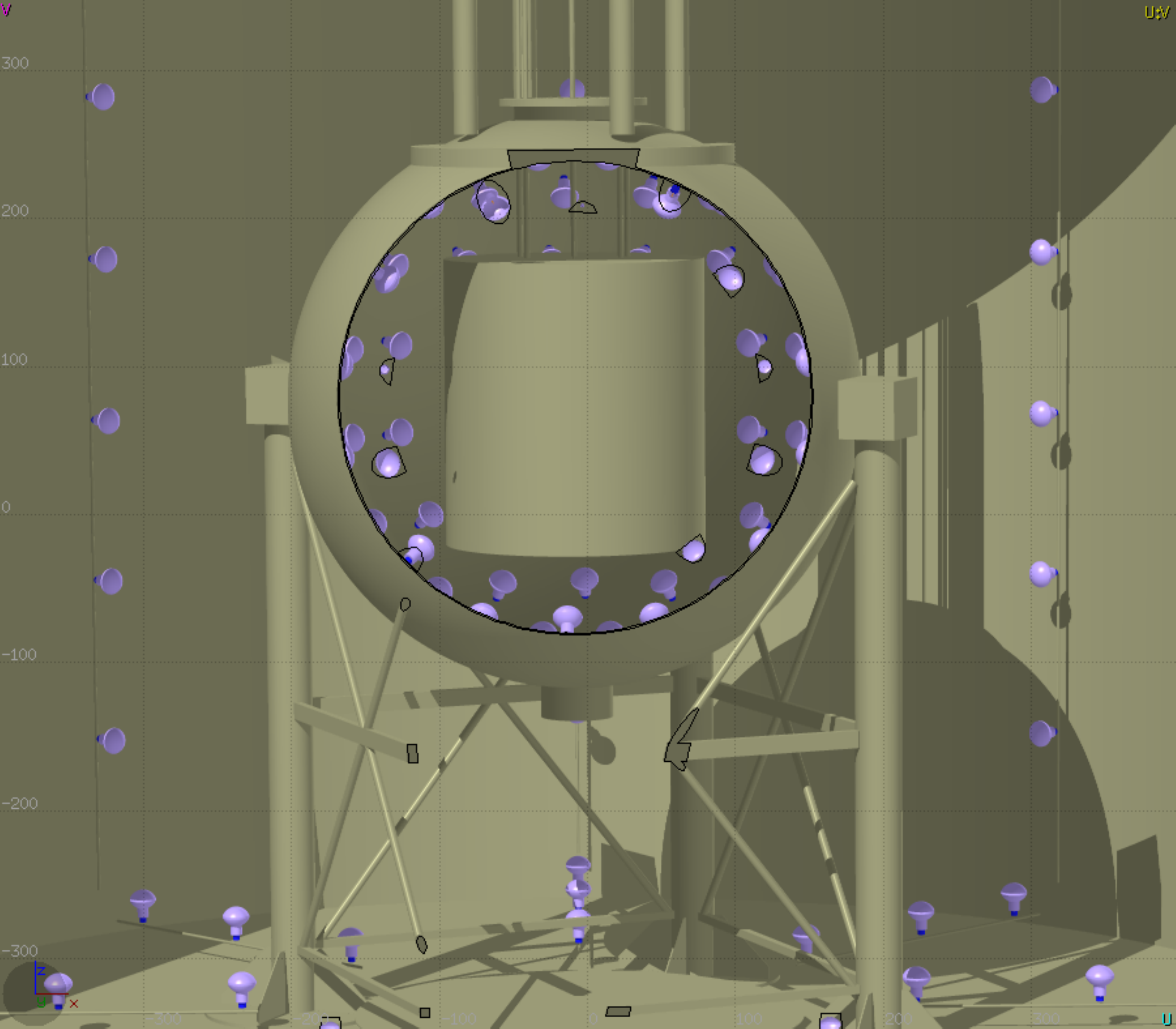}
    \includegraphics[width=.495\textwidth]{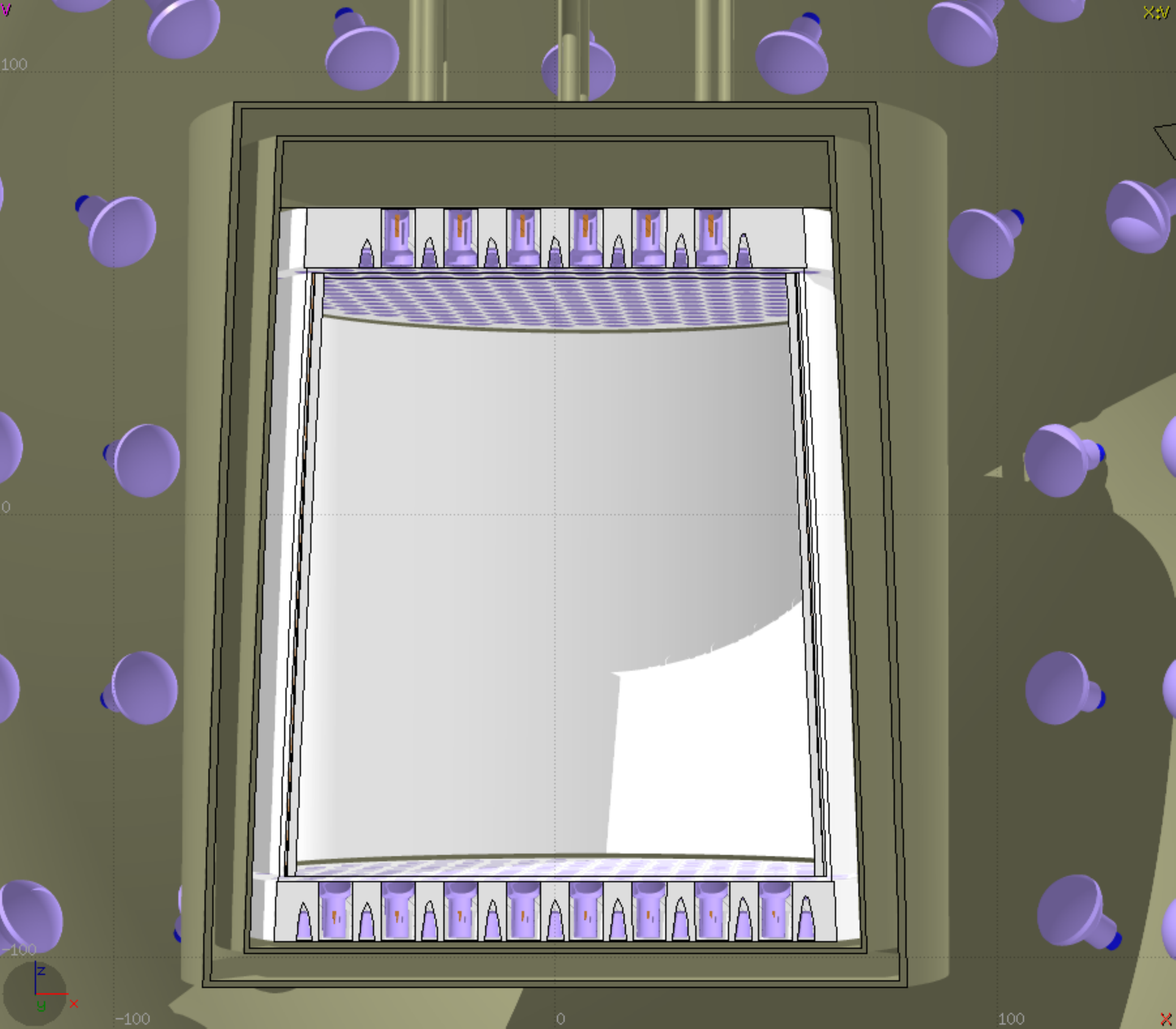}

    \caption{ FLUKA implementation of the \darkside experiment. 
              The Dewar containing the TPC is shown mounted inside
              the LSV on the left while the graphs on the right
              expose details of the geometry implemented for the
              TPC.   Top row: \Darkside-50; \,\,
                     Bottom row: \Darkside-G2.
            }
   \label{fig:outer}
   \end{figure}

 The \darkside detector consists of a cylindrical, two-phase underground
 liquid argon Time Projection Chamber (TPC)~\cite{ds50}.  The TPC is
 contained inside a thin-walled stainless steel Dewar.  The sensitive
 region is viewed by photomultiplier tubes from the top and bottom.  It
 is surrounded by a TPB\footnote{Tetraphenyl butadiene, fluorescent die}
 coated Teflon cylinder which acts as an optical
 reflector and wavelength shifter.  The Teflon cylinder also supports a
 set of copper rings which provide the electric field.  The \Darkside-50
 experiment with an active mass of 50 kg, is currently under commissioning.
 It features a sensitive volume of about 35 cm diameter and 35 cm height.
 The inner detector is immersed in a highly efficient, borated
 Liquid-Scintillator neutron Veto (LSV) to reduce external backgrounds and
 to monitor cosmogenic and radiogenic neutron backgrounds {\it in situ}.
 To further reduce cosmogenic backgrounds, the LSV is surrounded by ultra
 pure water inside a large cylindrical water tank (CTF) which functions
 as muon veto \v{C}erenkov detector.  The CTF also provides a passive
 shield against low energy external backgrounds.  The implementation
 of \darkside in FLUKA is shown in Figure~\ref{fig:outer}.

 In addition to its scientific reach, \darkside also serves as a prototype
 for the development of a ton-sized, next generation TPC. The present
 design of the veto detector system permits the upgrade to \Darkside-G2
 with a sensitive mass of approximately 3.3 tons using underground liquid
 argon depleted of the $^{39}$Ar isotope.
 The ton-sized TPC features a sensitive volume of about 150 cm diameter
 and 135 cm height.  In general, the rate of cosmogenic neutron background
 scales with the size of the sensitive detector volume and is
 substantially larger for \Darkside-G2 compared to that of \Darkside-50.
 Similarly, the increased size of the TPC displaces a larger amount
 of liquid scintillator inside the LSV and reduces the efficiency of
 detecting neutrons.

 \subsubsection{Counting Test Facility}

 The CTF~\cite{ctf} is modeled in this simulation as a cylindrical tank of
 11 meter diameter and 10 meter height constructed from 8 mm thick carbon
 steel.  It is filled with ultra pure water and placed on an additional
 10 cm thick layer of steel.  The inside tank surfaces are
 covered by sheets of a special type of layered Tyvek foil~\cite{tyvek}
 to enhance light collection.  For the relevant spectral range determined
 by the produced \v{C}erenkov light and the sensitivity of the PMTs, a
 reflectivity in water of greater than 95\% is expected.  However, in
 the simulation a
 more conservative constant reflectivity of 80\% was assumed.  \v{C}erenkov
 light produced in the water is measured by 80 ETL-9351 eight-inch PMTs which
 are placed along the floor and the walls of the CTF.  The production,
 propagation and detection of the \v{C}erenkov photons in the CTF is simulated
 in detail by FLUKA using a constant refractive index of
 1.334 and a realistic absorption spectrum for pure water \cite{pureH2O}.
 The photo sensors are modeled with their measured quantum efficiency
 spectrum scaled by a 70\% photoelectron collection efficiency
 \cite{McCarty}.  A study to evaluate the placement of the PMTs indicated
 only a weak dependence of the overall efficiency as a
 function of the spatial distribution of the optical sensors.   
 The insensitivity to the spatial dependence is attributed
 to the low absorption in pure water combined with the high surface
 reflectivity.

 \subsubsection{Liquid-Scintillator neutron Veto}

 The LSV \cite{lsv} is implemented as a 4 meter diameter stainless steel
 sphere with 8 mm wall thickness.  It is filled with a borated liquid
 scintillator consisting of a 1:1 mixture of Pseudocumene and
 Trimethylborate.  The sphere is located inside the CTF and both the
 inside and the outside surfaces are covered by the Tyvek reflector.
 The LSV is equipped with 110 low-background glass-bulb Hamamatsu
 R5912-HQE-LRI eight-inch  PMTs which are mounted on the sphere facing
 inward.  Optical processes inside the LSV were not simulated.   
 Instead, the un-quenched raw energy deposition per event is recorded.

 \subsubsection{Inner detector}

 Both, the DarkSide-50 and the DarkSide-G2~\cite{ds50}
 configurations of the inner
 detector were implemented in the FLUKA simulation.  Initial studies
 were carried out for the smaller DarkSide-50 design, however the
 following results are for the ton-sized DarkSide-G2 geometry.
 In this case, the cosmogenic background requirements are more
 stringent as DarkSide-G2 has a much larger sensitive volume.  Thus the
 results can be viewed as a conservative upper limit to those expected
 for DarkSide-50.  The geometry and material 
 composition of the Dewar and TPC for the  DarkSide-50 setup were
 modeled in detail according to the latest detector design drawings.
 This was then used to prepare a realistic, scaled model for the
 ton-sized configuration.  In the simulation the sensitive volume is
 viewed by  283 three-inch Hamamatsu, low-background R11065 PMTs
 positioned on the top and an equal number on the bottom of the TPC.
 As for the LSV, no simulation of the optical processes inside the TPC
 were undertaken.  The number of particles entering the sensitive
 volume, their type and the total raw energy deposited per event were
 recorded.

 \subsection{Simulation results}

 As described above, events which were ``frozen'' on the
 walls inside of the experimental Hall C at LNGS, were
 transported by FLUKA through the section of the cavern which
 contains the \Darkside-G2 experiment.  Results are presented based
 on a total number of simulated cosmogenic events corresponding to a
 lifetime of approximately 34 years.  The statistical uncertainty of
 the results is on the order of a few percent, and in any event, is
 smaller than the systematic uncertainties.   Events for which at least
 one particle reached the CTF water tank were recorded as a first step.  
 The predicted rate for these cosmogenic events at the outside of the CTF
 is approximately  3.45 events per minute.   For about 23\% of the events,
 the original cosmogenic muon does not reach the CTF water tank.

 In a second step of the
 simulation, the complete \Darkside-G2 detector setup was then exposed to
 all events which were recorded at the outside of the CTF.
 All physics processes were turned on making use of the
 FLUKA defaults setting PRECISIO(n).   In this step of the simulation
 \u{C}erenkov photons were created inside the CTF water tank.   However,
 because of CPU considerations they were not initially transported.
 The rate of cosmogenic events with at least one particle reaching the LSV
 is predicted to be approximately  0.30  events per minute.   This
 reduction in rate is the result of both the smaller size of the
 volume and the passive shielding of the water tank.
 Similarly, the rate of cosmogenic events with at least one particle
 reaching the sensitive region inside the TPC is predicted by FLUKA to
 be  0.07  events per minute. 
                               Reported rates are upper
 limits since no energy deposition in the respective detectors was
 required.  The expected cosmogenic event rates are summarized in
 Table~\ref{rates}.

  \begin{table}[thb]
   \begin{center}
    \caption{\label{rates} Expected cosmogenic event rates}
    \begin{tabular}{lc}
      just outside of  & cosmogenic event rate (per minute) \\
      \hline
      CTF      &  3.45 \\     
      LSV      &  0.30 \\
      sensitive liquid argon region & 0.07 \\
      \end{tabular}
    \end{center}
   \end{table}

\begin{figure}[htb]
    \centering
    \includegraphics[width=.600\textwidth]{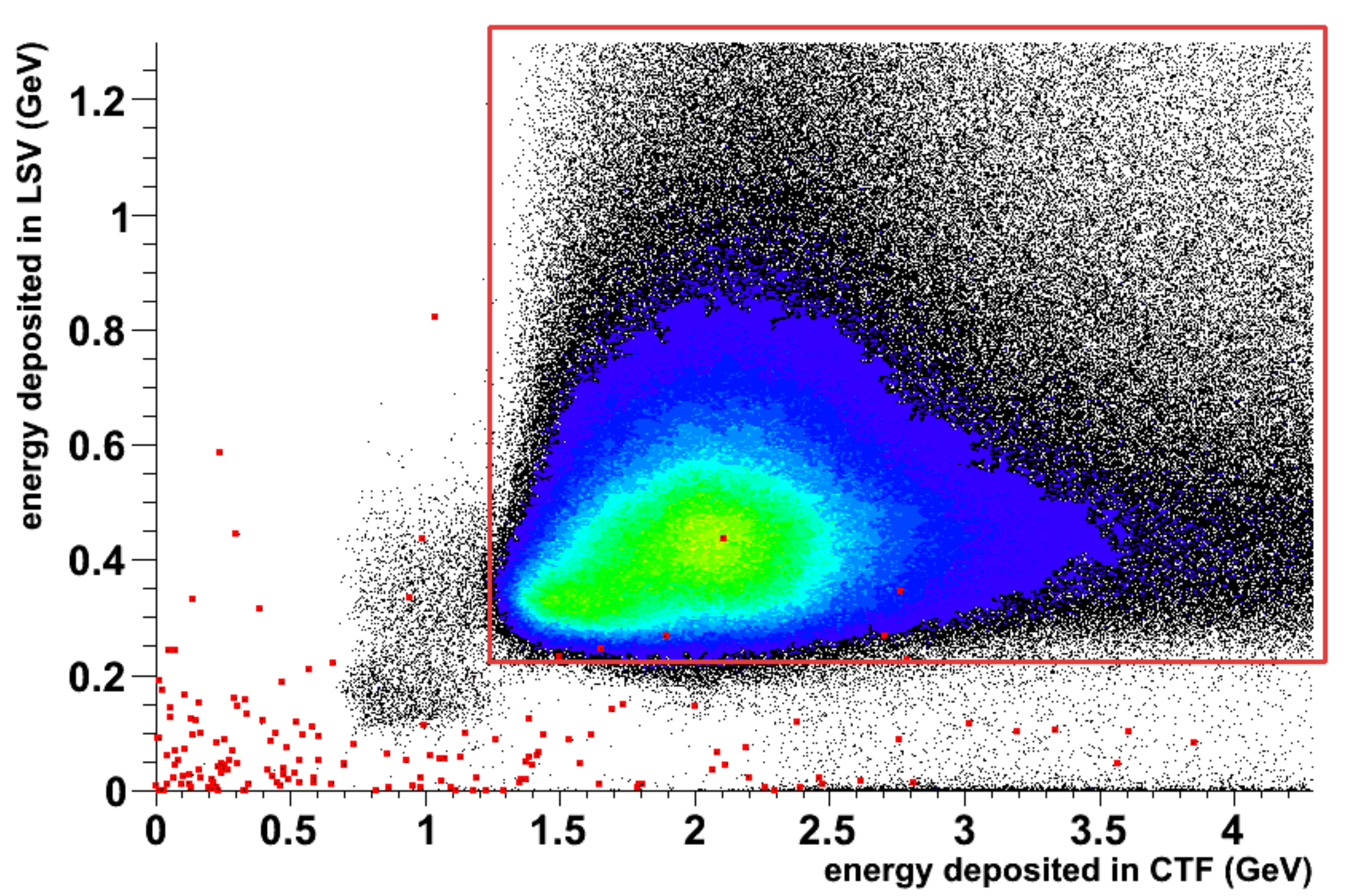}
    \caption{ Energy deposition inside CTF and LSV for cosmogenic
               events with at least one particle reaching the sensitive
                volume of the inner detector. }
   \label{fig:dEdE}
   \end{figure}

 In order to further study the predicted cosmogenic background for the
 \Darkside-G2 experiment, the subset of events with at least one particle
 reaching the inner sensitive region was considered.
%
 In Figure~\ref{fig:dEdE} the raw energy deposited inside the CTF and the LSV
 are graphed with respect to each other.   Cosmogenic events with at least
 one particle reaching the sensitive liquid argon volume almost always
 deposits  a significant amount of energy inside the veto detectors.
 The dominant region in the scatter plot is indicated by the red box and
 limited by  dE$_{(CTF)}$\,$>$\,1.3 GeV  and  dE$_{(LSV)}$\,$>$\,0.2 GeV. 

 This region corresponds to events with the original cosmogenic muon traversing
 both outer detectors.  The superimposed color contour plot indicates the shape
 of the most frequent energy deposition for cosmogenic events resulting from the
 ionization of the relativistic muons.  Events with similar energy deposited
 inside the CTF but with dE$_{(LSV)}$\,$<$\,0.2 GeV  indicate that 
 the original muon traversed the CTF but missed the LSV.  A smaller set of
 events in the region of  0.7 GeV\,$<$\,dE$_{(CTF)}$\,$<$\,1.3 GeV results
 from low energy cosmogenic muons which traverse the water tank but stop
 inside the LSV.

 The most difficult cosmogenic events to veto are found
 close to the origin of the graph.  For these events, little energy is
 deposited both inside the CTF and the LSV.  Cosmogenic events predicted by
 FLUKA without a direct muon into the CTF are superimposed on the graph with
 solid red symbols.  {    Approximately 7.9 of these events per year are             
 expected according to the simulation.}  Practically all cosmogenic events          
 which have small energy deposition in both the CTF and LSV fall
 into the class of events with no direct muon entering the CTF.

  \vskip 3mm

 The most important background events for the direct dark matter search
 consist of neutron-induced recoils in the sensitive volume from undetected
 neutrons.  In the next step of the simulation, all events for which at least
 one neutron (but $<$ 50 coincident particles~\footnote{
   This technical cut to reduce processing times only affects approximately
   4\% of the selected events. The suppressed events all deposit more than
   2 GeV of raw energy in the LSV as well as the CTF.
                                          }) reached the sensitive liquid
 argon region were reprocessed with full treatment of optical processes
 inside the CTF.  A total of 19735 of these events, or 581 events per year,
 are predicted by FLUKA.  Only the raw energy deposited inside the LSV and
 the sensitive region of the TPC are available in the current simulation.
 Therefore, conservative criteria were defined to select events which are
 considered detectable by the LSV:  dE$_{(LSV)}$\,$>$\,1 MeV, and for events
 which fall into the energy range of a possible dark matter signal
 in the TPC:  0\,$<$\,dE$_{(TPC)}$\,$<$\,1 MeV~\cite{dsg2}.

  \begin{figure}[htb]
    \centering
    \includegraphics[width=.495\textwidth]{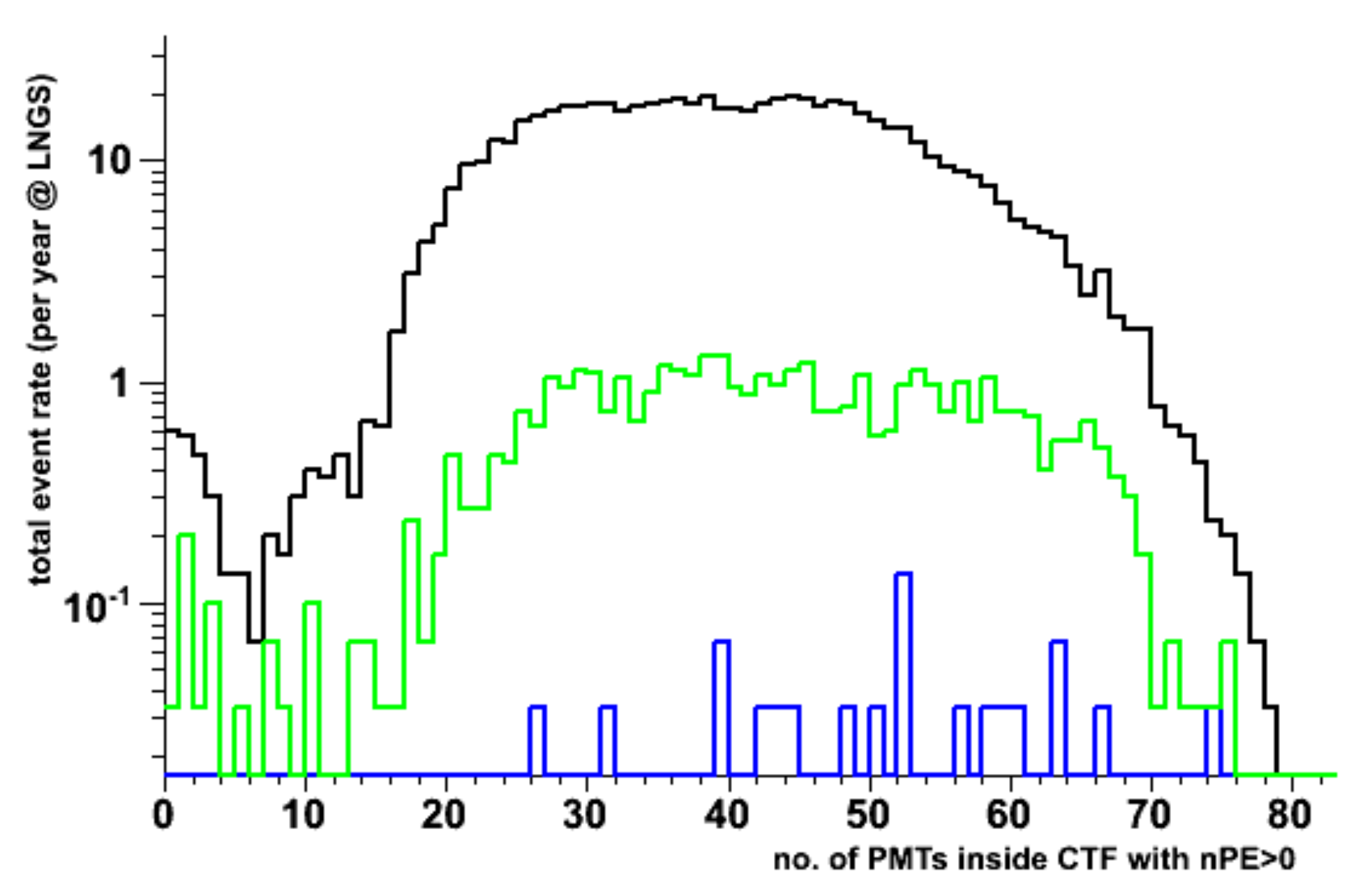}
    \includegraphics[width=.495\textwidth]{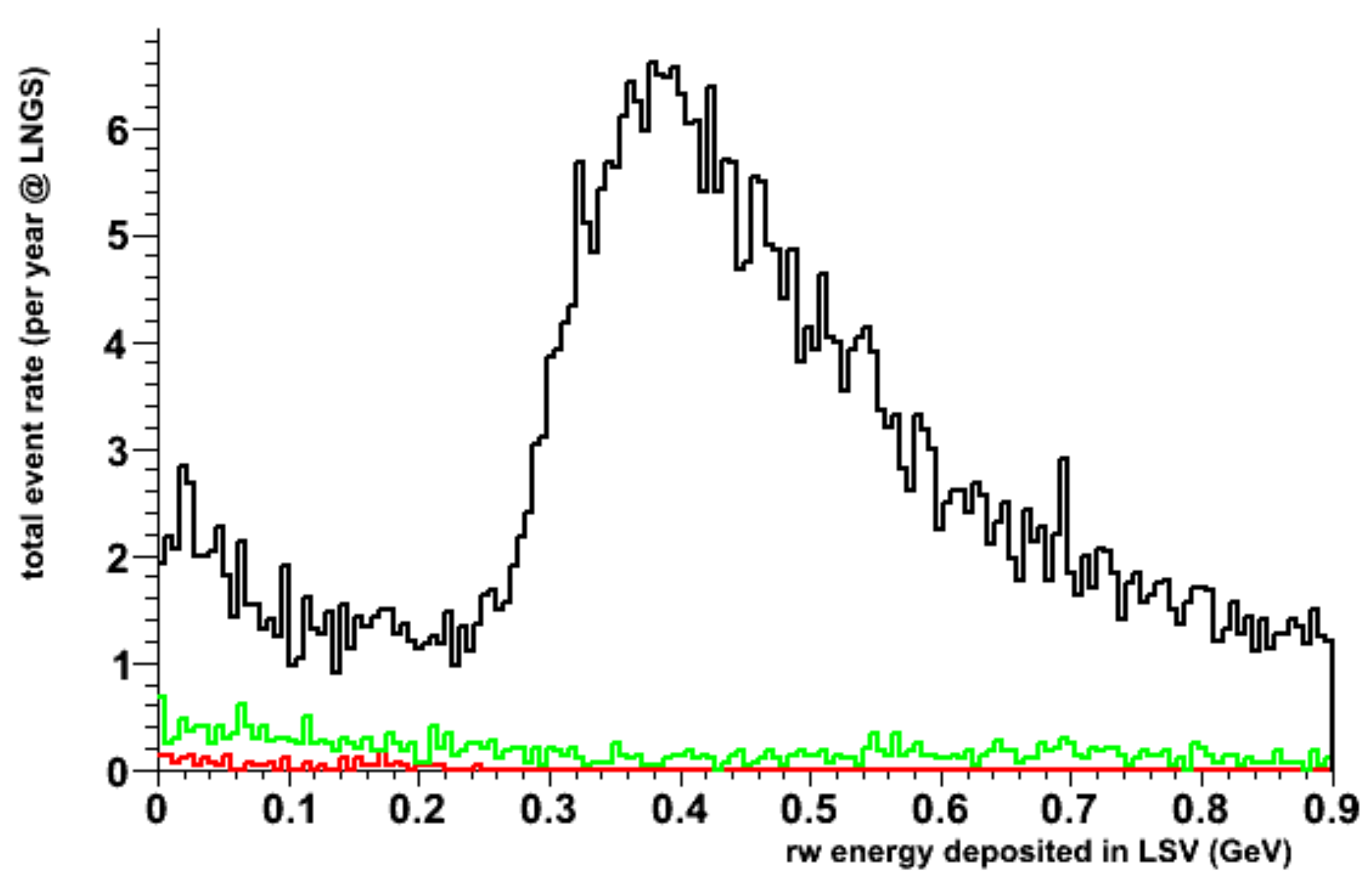}
    \caption{ (a) Number of PMTs with at least 1 registered photoelectron.
                  black: all events, green: raw dE inside TPC\,$>$\,0 and
                  $<$\,1 MeV, and blue: raw dE inside the LSV\,$<$\,1 MeV.
              \,\,
              (b) Un-quenched energy deposited inside the LSV. black:
                  all events, green: raw dE inside TPC $>$\,0 and $<$\,1 MeV,
                  and red: $<$\,10 PMTs with at least 1 registered
                  photoelectron.  
            }
   \label{fig:spectra}
   \end{figure}

 The predicted response of the veto detectors is shown in
 Figure~\ref{fig:spectra} for an equivalent lifetime of approximately
 34 years.  The number of events found for the CTF as a function
 of PMTs which register a signal (one or more photoelectrons) is given by
 the black histogram on the left.  The events shown by the blue histogram
 are found if the energy deposited inside the LSV is limited to less
 than 1 MeV.  The black histogram in the graph on the right shows the
 predicted energy spectrum for the LSV.  Limiting the sample to events
 with less than 10 PMTs which register a signal inside the CTF reduces the
 energy spectrum to the events shown by the red histogram.   The effect
 of selecting events in the energy range of interest for the TPC is
 indicated for both distributions by the green histograms.

\begin{figure}[hbt]
    \centering
    \includegraphics[width=.600\textwidth]{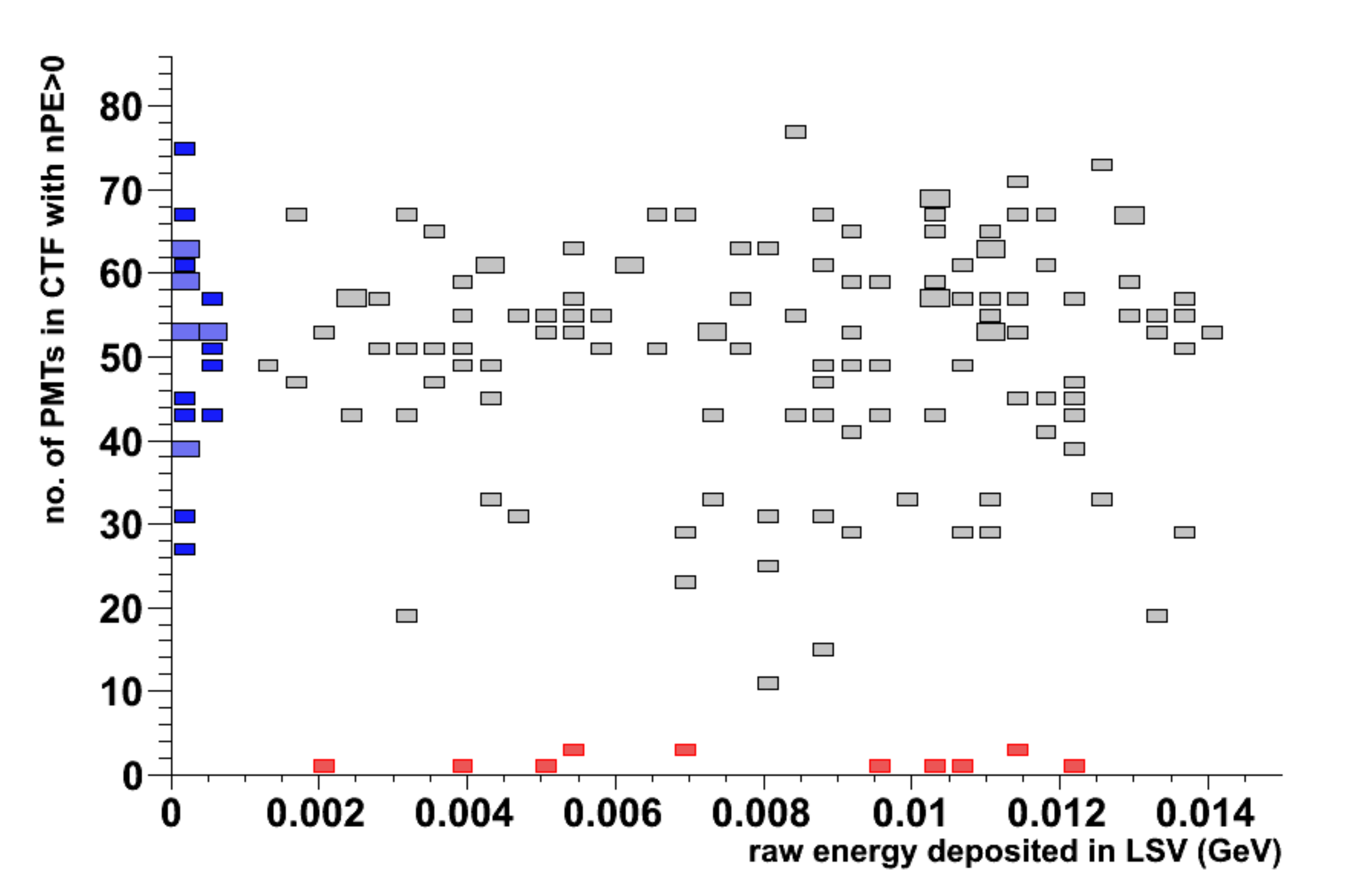}
    \caption{ Shown is the of number of PMTs which registered at least one
              photoelectron inside the CTF versus the un-quenched energy
              deposited in the LSV.  The red and blue color coded entries
              correspond to the respective events selected in
              Figure~\ref{fig:outer}.
            }
   \label{fig:2Dnew}
   \end{figure}

 The same information with focus on events with energy less than 14 MeV
 deposited in the LSV are shown in Figure~\ref{fig:2Dnew}.  The number of
 PMTs with a signal inside the CTF is graphed versus the raw energy
 deposited in the LSV.  Ten events with less than 3 PMTs recording a signal
 inside the CTF are considered to be missed by the muon veto and are
 colored red in the plot.  Similarly, twenty-one events with a raw energy
 deposition of less than 1 MeV inside the LSV are considered missed 
 and are colored blue in the graph.

 The FLUKA simulation predicts approximately 581 events per year in which at
 least one cosmogenic muon-induced neutron enters the sensitive liquid argon
 volume.  Out of these events, only 0.3 events per year fail to
 cause a signal in 3 or more PMTs of the CTF muon \u{C}erenkov veto.  At the
 same time, only 0.6 events per year deposit less than 1 MeV inside the
 LSV.  For a simulated live-time of approximately 34 years at LNGS, no event
 where a neutron reached the sensitive liquid argon region but
 failed to trigger at least one of the two veto detectors is observed.
 These reported rates do not take advantage of additional discrimination
 information from the TPC which can be used to further reject cosmogenic
 neutron background.

 The cosmogenic muon-induced neutron background rate to the \Darkside-50
 experiment is significantly smaller than for the much larger
 \Darkside-G2 configuration, while the same veto detectors are used
 for both implementations.  The rejection of external cosmogenic
 neutrons by the LSV increases for \Darkside-50 since there is more
 liquid-scintillator volume. In addition, this
 also increases the amount of passive shielding.

 The presented background evaluation was based on cosmogenic
 muon-induced secondaries which are prompt and thus can be vetoed by
 the outer detector systems.  The contribution from delayed neutrons
 after $\beta$-decay of muon-induced precursor nuclei, which are similar
 to detector internal radiogenic neutrons, are small as discussed
 in~\cite{delayed}.

 \section{Conclusions}

 FLUKA predictions for cosmogenic muon-induced neutron backgrounds
 are found to be in reasonable agreement with available
 data. A few additional problems were identified and will be addressed
 in future versions of the FLUKA code.  Corrections relating to the
 $(\gamma, n)$ reaction on $^{12}C$ for example,
 can be approximately included in the final results, as was discussed
 in the text.  However, the discrepancy and suggested correction
 reported in Ref.~\cite{hime} were found invalid.  On the other
 hand, a compelling analysis of the underpredicted single neutron
 multiplicity yield pointed to a potential problem in the missing 
 contribution of low energy, electro-produced neutrons in FLUKA.

 The need of a detailed description of the full muon-induced
 radiation field for the considered underground site was described and
 shown to be important. Steps to explicitly prepare the cosmogenic
 radiation field with the FLUKA simulation package were described and
 the muon-induced radiation field which was
 used to carry out benchmark studies, was then applied to the
 \darkside experiment. Background predictions for a direct dark
 matter search experiment were obtained for the next generation,
 ton-sized two-phase underground liquid argon detector.  It was
 found that the proposed dual active-veto system for the experiment
 provides sufficient shielding against cosmic radiation at the LNGS
 depth for a ton-sized \Darkside-G2 experiment for more than 5 years.

 Background levels for the \Darkside-50 experiment, which makes
 use of the same veto detector system, are expected to be significantly
 reduced because of the smaller size and the consequently increased
 volume of the liquid-scintillator shield and thus adequate for the
 smaller detector.  In the future it is essential to monitor
 the cosmogenic neutron background levels, with focus on the veto
 detector system, in order to continue to benchmark the simulation.

\acknowledgments

 This work was supported in part by NSF awards 1004051 and 1242471.
 We would like to thank Alfredo Ferrari and
 Maximilliano Sioli for FLUKA related help and acknowledge the use of
 the FLUKA graphical user interface FLAIR.  We also would like to thank
 Davide D'Angelo, Quirin Meindl and
 Michael Wurm of the Borexino collaboration.
 Finally, we like to thank Vitaly Kudryavtsev for lively discussions,
 in particular about early measurements of the muon-induced neutron yield.

\bibliographystyle{JHEP}
\bibliography{references}

\end{document}